\documentclass[aps,prd,twocolumn,nofootinbib,superscriptaddress]{revtex4-1}
\usepackage[utf8]{inputenc}
\usepackage{graphicx}% Include figure files
\usepackage{subfigure}
\usepackage{datatool}
\usepackage{amsmath,amssymb}
\usepackage[scientific-notation=true]{siunitx}
\usepackage{lineno}
\usepackage{natbib}
\usepackage[dvipsnames]{xcolor}
\usepackage{xspace}
\usepackage{orcidlink}

\DTLsetseparator{,}

\def\gc{Galactic Center\xspace}
\def\gce{Galactic-Center excess\xspace}
\def\gch{Galactic-Center\xspace}
\def\gw{gravitational wave\xspace}

\def\gwh{gravitational-wave\xspace}
\def\gws{gravitational waves\xspace}

\def\dmh{dark-matter\xspace}

\def\msps{millisecond pulsars\xspace}
\def\msp{millisecond pulsar\xspace}
\def\lvk{LIGO/Virgo/KAGRA\xspace}

\newcommand{\TFFT}{T_\text{FFT}}

\newcommand{\Izz}{I_{\rm zz}}
\newcommand{\Bint}{B_{\rm int}}
\newcommand{\Bext}{B_{\rm ext}}

\newcommand{\bea}{\begin{eqnarray}}
\newcommand{\eea}{\end{eqnarray}}
\newcommand{\be}{\begin{equation}}
\newcommand{\ee}{\end{equation}}

\newcommand{\pgw}{P_{\rm GW}}

\newcommand{\fermi}{\textit{Fermi}\xspace}

\begin{document}

\title{Probing the pulsar explanation of the Galactic-Center GeV excess using continuous gravitational-wave searches }

\author{Andrew L. Miller\,\orcidlink{0000-0002-4890-7627}}
\email{andrew.miller@nikhef.nl}
\affiliation{Université catholique de Louvain, B-1348 Louvain-la-Neuve, Belgium}
\affiliation{Nikhef -- National Institute for Subatomic Physics,
Science Park 105, 1098 XG Amsterdam, The Netherlands}
\affiliation{Institute for Gravitational and Subatomic Physics (GRASP),
Utrecht University, Princetonplein 1, 3584 CC Utrecht, The Netherlands}

\author{Yue Zhao}
\email{zhaoyue@physics.utah.edu}
\affiliation{Department of Physics and Astronomy, University of Utah, Salt Lake City, UT 84112, USA}

\begin{abstract}
Over ten years ago, \fermi observed an excess of GeV gamma rays from the Galactic Center whose origin is still under debate. One explanation for this excess involves annihilating dark matter; another requires an unresolved population of millisecond pulsars concentrated at the Galactic Center. In this work, we use the results from LIGO/Virgo's most recent all-sky search for quasi-monochromatic, persistent gravitational-wave signals from isolated neutron stars, which is estimated to be about 20-50\% of the population, to determine whether unresolved millisecond pulsars could actually explain this excess. First, we choose a luminosity function that determines the number of millisecond pulsars required to explain the observed excess. 
Then, we consider two models for deformations on millisecond pulsars to determine their ellipticity distributions, which are directly related to their gravitational-wave radiation.
Lastly, based on null results from the O3 Frequency-Hough all-sky search for continuous gravitational waves, we find that a large set of the parameter space in the pulsar luminosity function can be excluded. We also evaluate how these exclusion regions may change with respect to various model choices. Our results are the first of their kind and represent a bridge between gamma-ray astrophysics, gravitational-wave astronomy, and dark-matter physics. 
\end{abstract}

\maketitle
\section{Introduction}\label{sec:intro}

A tantalizing excess of GeV gamma rays was observed by \fermi over 10 years ago coming from the \gc, and yet its origin has remained elusive. While early studies suggested that the almost spherically symmetric \gce spatial morphology was well-fit by \dmh models \cite{Goodenough:2009gk,Hooper:2010mq,Hooper:2011ti,Gordon:2013vta,Daylan:2014rsa,Calore:2014nla,Abazajian:2014fta}, and that recent cosmic-ray burst events in our galaxy
could also explain this excess \cite{Petrovic:2014uda,Carlson:2014cwa,Cholis:2015dea}, an astrophysical explanation of unresolved \msps also appears consistent with the observed excess \cite{Abazajian:2010zy,Calore:2014xka,Yuan:2014rca,Petrovic:2014xra,Ye:2022yxt}. The debate between these two explanations is intense: some studies claim that the spatial morphology of the \gce matches better with the mass distribution of the galactic bulge  \cite{Macias:2016nev,Bartels:2017vsx,macias:2019omb,Pohl:2022nnd}, while other studies indicate a preference for a spherically symmetric distribution \cite{DiMauro:2021raz,Cholis:2021rpp}. There have also been potential detections of gamma rays from point sources in the inner Galaxy \cite{Lee:2014mza,Lee:2015fea,Buschmann:2020adf}, but it was shown recently that systematic biases favoring individual sources may exist in these works \cite{Leane:2019xiy,Leane:2020nmi,Leane:2020pfc}. Furthermore, predictions for the fraction of the \gce explained by \msps range from a few percent \cite{Hooper:2013nhl,Hooper:2016rap} to 100\% \cite{Ploeg:2020jeh}, depending on the luminosity function chosen.%, which indicates the present difficulties in explaining this excess.

The general theme behind these studies is to devise a luminosity function to fit the observed \gce, based on minimal assumptions \cite{Bartels:2017vsx,List:2021aer} or astrophysics, e.g. assuming luminosity functions identical to those from known \msps in globular clusters \cite{Hooper:2016rap}, asserting that accretion-induced collapse is responsible for creating a \msp population \cite{Gautam:2021wqn}, or allowing emissions from low-mass X-ray binaries to compose the \gce \cite{Cholis:2014lta,Haggard:2017lyq}. However, since these choices are required to generate the same observed excess, only some studies of {X-rays \cite{Berteaud:2020zef}, TeV gamma-rays \cite{Macias:2021boz} and radio observations \cite{Calore:2015bsx} have allowed to actually exclude luminosity functions}.%, nor determine consistently what fraction of the \gce could be caused by \msps. % \yz{Some clarification here.}

Another approach is thus needed to test the viability of the \msp hypothesis.
%one that does not rely on whether \msps need to emit individually or ensemble, and one that can actually rule out portions of the luminosity function parameter space.
% nor on the total number of \msps necessary to explain the \gce \cite{Dinsmore:2021nip}, nor on uncertain luminosity functions for the \gce.  \yz{We do need these information as an input to predict the GW radiation. I am a little bit confused about this comment.}
% \com{I meant that even though our method takes as input this information, we don't actually hav to explain the GeV excess - we just use them to constrain the GeV excess. }
In this paper, we show that all-sky searches for continuous waves, i.e. quasi-monochromatic, persistent signals from isolated, asymmetrically rotating neutron stars concentrated around the \gc, 
can constrain the \msp hypothesis for a chosen luminosity function 
% \yz{[This demonstrates that we still have the luminosity dependence.]} 
and provide a complementary probe of the \gce. 

Gravitational waves could be emitted by neutron stars with deformations on their surfaces, which would then cause a ``spin-down'', a decrease in the rotational frequency of the star over time, of $\mathcal{O}(<10^{-9})$ Hz/s. The size of this deformation is a priori unknown and could vary amongst neutron stars \cite{Woan:2018tey}; however,
%though some evidence indicates that a minimum does exist for \msps \cite{Woan:2018tey}. 
% While \gwh signals originating from \msps would be weaker than the electromagnetic ones, 
% advanced LIGO \cite{2015CQGra..32g4001L}, Virgo \cite{acernese2014advanced} and KAGRA \cite{aso2013interferometer} can measure extremely small displacements induced by a passing \gw, independently of any gamma-ray emission from pulsars. This means that LIGO/Virgo/KAGRA could detect \gws from sources that \fermi does not see, i.e. those that have small electromagnetic luminosities. 
contrary to \fermi, advanced LIGO \cite{2015CQGra..32g4001L}, Virgo \cite{acernese2014advanced} and KAGRA \cite{aso2013interferometer} do not rely on electromagnetic emissions, meaning they could potentially detect \gws from sources \fermi may not see.

In fact, the strength of a superposition of signals from a stochastic \gwh background of isolated \msps in the Galactic Center was calculated in \cite{Calore:2018sbp}, assuming a fixed ellipticity and moment of inertia for the population. 
% and a sensitivity estimate in future detectors was provided, indicating the number of pulsars that could explain this excess based on \fermi measurements. 
However, the authors did not systematically try to exclude the luminosity function, nor test robustness of their calculations against different modeling assumptions for the \msp population. Furthermore, a stochastic \gwh background was actually searched for in the \gc recently \cite{Agarwal:2022lvk}, resulting in constraints on the ellipticity of \msps there; however, this search was conducted independently of any knowledge of the GeV excess. In contrast, here we consider \msps emitting individually-detectable signals, directly link \gwh searches and the observed luminosity of the GeV excess, and put data-driven constraints, while testing the robustness of our modeling choices.
% Furthermore, recent \gwh searches have surpassed the so-called ``spin-down limit'' for \msps \cite{LIGOScientific:2020gml}, indicating that they would be sensitive to the case in which \gws take less than 100\% of the total rotational energy of neutron stars.  \yz{Not sure how the last sentence fits the rest of this paragraph.}

To address the \gce problem using \gws, we will use results from one of the most recent all-sky searches of the latest \lvk data, O3, that targeted isolated neutron stars with \gwh frequencies between [10, 2048] Hz and spin-downs between $[-10^{-9},+10^{-8}$] Hz/s \cite{LIGOScientific:2022pjk} with the Frequency-Hough method \cite{Astone:2014esa}. We will specifically consider the portion of the sky that contains the \gc. Additionally, we choose to use upper limits from searches for isolated neutron stars because these searches can reach the \gc ($\sim$ 8 kpc), which cannot yet be reached by all-sky searches for neutron stars in binary systems\footnote{{{The O3a all-sky Binary SkyHough search \cite{LIGOScientific:2020qhb} only analyzed \gwh frequencies up to $300$ Hz, and did not consider a frequency derivative term in the \gwh waveform, which limits the search sensitivity to, at best, $1-2$ kpc for $\epsilon=10^{-4}$, and to much lower distances for smaller ellipticities, which are expected for \msps.}} }, thus our results only constrain isolated \msps that could comprise $\sim 20\%$ of the population \cite{Jiang:2019hal}, and some binary systems with particular orbital parameters-- see App. \ref{app:mspbin}.  %Though no \gws were detected, we can still use this result to exclude portions of the luminosity function parameter space, which effectively determines a maximum number of \msps that could explain the \gce.

\section{Millisecond pulsars}

\subsection{Gravitational-wave emission}
% - ATNF catalog

Gravitational waves from an isolated neutron star could be emitted due to a deviation from axial symmetry, which can be written in terms of a dimensionless equatorial ellipticity $\epsilon$, defined in terms of the star's principal moments of inertia \cite{maggiore2008gravitational} $ \epsilon \equiv {|I_{\rm xx} - I_{\rm yy}|}/{I_{\rm zz}}$. The value of
$\Izz$ is at $\mathcal{O}(10^{38}-10^{39})$ kg$\cdot$m$^2$, depending on the unknown neutron star equation of state \cite{Steiner:2014pda,Breu:2016ufb}. In this study, we choose three representative values $(10^{38},5\times 10^{38},10^{39})$kg m$^2$.
% \yz{Show typical values of $I_{\rm zz}$, and mention that we will take two benchmark numbers in our study.} 
The \gwh amplitude $h_0$ is directly proportional to the ellipticity \cite{maggiore2008gravitational}:
\begin{equation}\label{eq:h0}
    h_0 = \frac{16\pi^2 G}{c^4} \frac{I_{\rm zz} \epsilon f_{\rm rot}^2 }{d},
\end{equation}
where $d$ is the star's distance from the Earth, $f_{\rm rot}$ is the star's rotational frequency.% $G$ is Newton's gravitational constant, and $c$ is the speed of light.

We plot a distribution of \gwh frequencies given by the Australian Telescope National Facility (ATNF) catalog \cite{Manchester:2004bp} in App. \ref{app:dist}.
% we plot the \gwh frequencies $f_{\rm GW}=2f_{\rm rot}$ that correspond to potential \msps, i.e. $f_{\rm rot}>60$ Hz. 
In this study, we adopt the strategy in \cite{DeLillo:2022blw} and assume this distribution of frequencies is representative of the unknown rotational frequencies of \msps.

It is also useful to introduce the spin-down limit ellipticity $\epsilon^{\rm sd}$. This is the maximum allowed ellipticity of a neutron star, assuming that all of the rotational energy lost by a \msp is converted into \gws \cite{LIGOScientific:2020lkw}:
%  \begin{equation}\label{eq:h0sd}
%      h_{0}^{\rm sd}=\frac{1}{d}\left(\frac{5GI_{\rm zz}}{2c^3}\frac{|\dot{f}_{\rm rot}|}{f_{\rm rot}}\right)^{1/2},
%  \end{equation}
%  where $\dot{f}_{\rm rot}$ is the spin-down of the \msp. 
% By equating Eq. (\ref{eq:h0}) and Eq. (\ref{eq:h0sd}), we derive the ellipticity at the spin-down limit $\epsilon^{\rm sd}$:
 \begin{equation}
     \epsilon^{\rm sd}=\sqrt{\frac{5c^5}{2G}}\frac{1}{16\pi^2}\sqrt{\frac{ |\dot{f}_{\rm rot}|}{\Izz f_{\rm rot}^5}}.
 \end{equation}

%The ellipticity is the primary driver for the amount of energy released in \gws from deformed \msps. Therefore, 
In the following study, we consider two models to determine the ellipticity distribution of \msps. The first is to model the deformation as being caused by a strong internal magnetic field misaligned with the star's rotational axis \cite{Lander:2011yr,Mastrano:2011aa,Lander:2013oea}. The second is to require that the \gwh radiation of \msps is a fixed fraction of the total energy loss \cite{Ushomirsky:2000ax,Horowitz:2009ya}. %, i.e. at an ellipticity smaller than $\epsilon^{\rm sd}$ .

% In this work, we assume the spin-down limit when calculating the ellipticity probability density distribution in terms of crustal strain braking (see section \ref{subsec:crust}), though we consider that at most 10\% of rotational energy goes into \gws, in order to set conservative constraints. 

\subsection{Deformation caused by magnetic field}\label{mag-def}

Neutron stars cannot remain spherical in the presence of a strong \emph{internal} magnetic field, $\Bint$ \cite{Maggiore:2018sht,Chandrasekhar:1953zz}. If this field does not align with the rotational axis, it could sustain a deformation on the surface. Specifically, assuming a superconducting core, 
the ellipticity is related to $\Bint$ as \cite{LIGOScientific:2020gml},
$\epsilon \approx 10^{-8} \left(\frac{B_{\rm int}}{10^{12} \, \rm Gs}\right) $.

The internal magnetic field of a pulsar is not directly observable. We have to derive its value using the pulsar's measured external magnetic field, $\Bext$. %The connection between $\Bext$ and $\Bint$ is not well understood. 
In this study, we consider a range of ratios when calculating the probability of \gwh detection based on O3 search results, but we take a benchmark value as $\Bint=150\Bext$, motivated by \cite{Ciolfi:2013dta,Haskell:2015psa}, when applying our results to exclude portions of the luminosity function parameter space (see Sec. \ref{sec:meth} and Sec. \ref{sec:results}). We also assume \msps near the \gc follow the same $\Bext$ distribution given in the ATNF catalog \cite{Manchester:2004bp}. However, the internal magnetic field could even be $10^4$ times larger than the external one \cite{Vigelius:2009eg,Priymak:2011zv}. 
% \yz{[Do you mean $\Bext\gg\Bint$ or the coefficient between $\Bext$ and $\Bint$ can be much larger than 100?]} 
% if the field is buried during a past phase of accretion and remains stable during the star's lifetime \cite{Mukherjee:2017rwl}. 
% Larger $\Bint$ would correspond to bigger ellipticities, which would improve the exclusion regions presented in this work, so $\Bint=150\Bext$ is a relatively conservative choice with respect to the range of possible ratios.

In App. \ref{app:dist}, we show the probability distribution of ellipticity, assuming an external magnetic field distribution reported in the ATNF catalog. %The peak of this distribution is $\epsilon\sim 3.16\times 10^{-10}$, corresponding to an external magnetic field of $\Bext\sim 3.16\times 10^{8}$ G. 

\subsection{Fixed fraction of \gwh energy loss} \label{subsec:crust}

Rather than relying on specific models to estimate the ellipticity distribution, we also employ ellipticities that are inferred from the ATNF catalog at $\sim 5\%$ or $\sim 10\%$ of the spin-down limit. %, i.e. 0.25\% or 1\% of the star's rotational energy is lost via \gws. 
%Choosing \gwh emission to occur at these levels could 
Such a choice is motivated by recent constraints on the \gwh equatorial ellipticities of \msps, in which the spin-down limit has been slightly surpassed for some known \msps \cite{LIGOScientific:2020gml}. We note that our assumed values are smaller with respect to the constraints on \msps in \cite{LIGOScientific:2020gml} %, which indicate that at most $\sim 50\%$ of rotational energy could be emitted via \gws. 
% and noting that if we observe these pulsars electromagnetically, it is likely that most of their rotational energy is emitted via electromagnetic radiation, not through \gws. 
This probability density distribution is given in App. \ref{app:dist}. 
% As a point of comparison, if $u=0.1$, $\epsilon\approx 10^{-6}$, which would be a huge deformation, relative to the ellipticity probability density function in figure \ref{fig:pdf_crust} that peaks at $\epsilon<10^{-10}$.
%Employing the probability density functions for the ellipticity in Fig. \ref{fig:ellip_pdf_I_1e38_factor_100} and Fig. \ref{fig:pdf_crust} will allow us to obtain conservative constraints on the role of \msps in explaining the GeV excess.

\subsection{Luminosity function for galactic-center GeV emission}\label{lumin-func}

The luminosity function is directly related to the number of \msps needed to explain the observed GeV excess, and is one of the dominant contributions to astrophysical uncertainties. In this paper, we use two well accepted benchmarks. The first is a luminosity function following a log-normal distribution ~\cite{Hooper:2016rap}:
\begin{align}
    \frac{dN}{dL} \propto \frac{dP(L)}{dL}= \frac{\log_{10} e}{\sigma_L \sqrt{2\pi} L}\exp{\left(-\frac{{\log_{10}^2 (L/L_0) }}{2\sigma_L^2}\right)},
    \label{eqn:log-normal}
\end{align}
% \begin{equation}
%     \frac{dN}{dL} \propto P(L)= \frac{\log_{10} e}{\sigma_L \sqrt{2\pi} L}\exp{\left(-\frac{{(\log_{10} L - \log_{10} L_0})^2}{2\sigma_L^2}\right)},
%     \label{eqn:log-normal}
% \end{equation}
where $L$ is the luminosity, and $L_0$ and $\sigma_L$ are two free parameters. Details on the ranges of these parameters are given in App. \ref{app:lum}. 

The second benchmark has a general power-law dependence on energy cut ($E_{\rm cut}$), magnetic field ($B$) and the spin-down power ($\dot E$) \cite{Ploeg:2020jeh}:
\begin{align}
    L\frac{dP(L)}{dL}= \eta E_{\rm cut}^{a_\gamma} B^{b_\gamma} \dot E^ {d_\gamma}.
    \label{eqn:power-law}
\end{align}
More detailed discussion of this luminosity function can be found in App. \ref{app:lum}.

%A detailed study has been performed in \cite{Ploeg:2020jeh}, and various model parameters are obtained by fitting to the observed GeV excess. To demonstrate how a GW search can be complementary, we take one benchmark model with $log_{10}(B_{\rm med}/G)=8.21$, $\sigma_B=0.21$, $log_{10}(\eta_{\rm med})=12$
%\begin{align}
%   p(log_{10}(x)|x_{\rm med},\sigma_x)\\
%   =\frac{1}{\sqrt{2\pi}\sigma_x}exp\bigg (-\frac{log_{10}(x)- log_{10}(x_{\rm med}))^2}{2\sigma_x^2}\bigg )
%    \label{eqn:power-law}
%\end{align}

We note that our results can easily be generalized to different luminosity functions, since the result from the \gwh search simply translates to a constraint on the total number of \msps detectable via \gws, i.e. $N_{\rm GW}$. For a different choice of the pulsar luminosity function, one simply needs to compare $N_{\rm GW}$ with the number of pulsars needed to explain the GeV excess to obtain a constraint on the luminosity function parameters.

Following the discussion in \cite{Dinsmore:2021nip}, we first calculate the total luminosity contributed by millisecond pulsars in the \gc: 
\begin{align}
     L_\text{GCE} = N_\text{MSP}\int_{L_\text{min}}^\infty L P(L) dL.
     \label{eqn:lgce}
\end{align}
Here $N_\text{MSP}$ is an overall normalization parameter, characterizing the number of millisecond pulsars, and $L_{\rm min}$ is the minimum detectable luminosity by \fermi. For $L_{\rm GCE}\approx 10^{37}$ erg/s (see App. \ref{app:lum} for how we obtain this number), we compute $N_{\rm MSP}$ for various choices of $L_0$ and $\sigma_L$, which will influence the number of detectable \gwh sources in O3.

% Later, $N_{\rm MSP}$ will be tested based on the results from the Frequency-Hough all-sky search for continuous waves in O3, see Sec. \ref{sec:meth}.

% Later, $N_{\rm MSP}$ will be tested based on the continuous \gwh results of the Frequency-Hough all-sky search in O3, see section \ref{sec:meth}.

\section{Method}\label{sec:meth}

A search for quasi-monochromatic \gwh signals originating from anywhere in the sky was performed using data from the third observing run of advanced \lvk \cite{LIGOScientific:2022pjk}. One algorithm, the Frequency-Hough \cite{Astone:2014esa}, tracks linear frequency evolution over time by mapping points in the time-frequency plane of the detector to lines in the frequency/frequency derivative plane of the source \cite{Astone:2014esa,Piccinni:2018akm}. Though all outliers were vetoed, competitive upper limits were set on the degree of deformation that neutron stars could have -- see App. \ref{app:all-sky_ul}.

In this study, we apply the results obtained in the Frequency-Hough all-sky search to calculate the number of detectable \msps at the \gc. We note that this search and the one that specifically targets a single sky pixel that completely covers the \gc \cite{LIGOScientific:2022lsr} are complementary to each other, and we will comment on the future optimization later. 
The \gch search can obtain a better sensitivity than the all-sky one, since it only looks at one pixel, significantly reducing the computational cost of the search, and can therefore use longer Fourier Transforms to look for quasi-monochromatic signals. Here, however, we would like to consider a larger spatial extent than that covered in the \gch search, i.e. greater than 150 pc from the \gc, which is why we use the all-sky search results. We therefore apply a correction factor to ``specialize'' the all-sky search results to the \gc, as described in App. \ref{app:all-sky_ul}.

% However, any dependence on sky position \emph{within} the \gc is lost in this search, which is why we focus on the all-sky search results here, where we want to be able to understand how the number of \msps would change as a function of location within the \gc, i.e. a larger extent than that which is considered in the \gch search.
% \yz{Can we briefly compare these two searches? Particularly the difference in analysis techniques?} 

In our study, we assume particular ellipticity distributions described in Sec. \ref{mag-def} and \ref{subsec:crust}. Furthermore, we apply the rotation frequency distribution measured in the ATNF catalog \cite{Manchester:2004bp} to determine the \gwh frequency from these \msps, as shown in App. \ref{app:dist}. We provide details on the conversion from the direct output of the \gwh search, $h_0$, to the ellipticity $\epsilon$ in App. \ref{app:all-sky_ul}. 
% \yz{May remove the latter half of this paragraph if we run out of space.} \yz{Also, the LVK result is about strain, not ellipticity. Here we may want to explicitly refer to the app for the conversion. The app spent a lot of time talking about all-sky to GC search, but not $h$ to $\epsilon$. We may want to add a little bit more of details there.} 

Let us calculate the probability for a \msp to be detectable through \gwh measurements, $\pgw$. The \gwh search leads to upper limits of the ellipticity as a function of the frequency, i.e. $\epsilon_{\rm UL}(f)$, at a given confidence level (here, 95\%). These limits mean that if there is one \msp whose rotation frequency is $f_{\rm rot}$ and ellipticity is larger than $\epsilon_{\rm UL}(f)$, it should have been detected by the search.
% \yz{We may add one sentence description on what this upper limit means, e.g. If there is one msp whose rotation frequency is $f$ and ellipticity is larger than $\epsilon_{\rm UL}(f)$, it should have been detected by LVK. } 
As an illustration, we present the results for the Frequency-Hough all-sky search in App. \ref{app:all-sky_ul}. 
% we present the result targeted for the \gch neutron star search in App. \ref{app:all-sky_ul}. 
From these upper limits in App. \ref{app:all-sky_ul}, we first calculate the probability that a neutron star has an ellipticity above the minimum detectable one for a given frequency. This is obtained by integrating the ellipticity distribution, given in Fig. \ref{fig:ellip_pdf_I_1e38_factor_100} or \ref{fig:pdf_crust}, over the value above $\epsilon_{\rm UL}(f)$.
% \yz{I do not understand this comment.}\com{it's meant to reinforce that elliptiicty and frequency are independent, and that the way in which I've written $\epsilon(f)$ only means that we are calculating the ellipticity at each frequency, not that the ellipticity depends on frequency.} 
We then integrate this quantity over the frequency distribution given in Fig. \ref{fig:f_pdf}. 
This gives
\begin{align}
P_{\rm GW}&=& \int_{\log_{10} f_{\rm min}}^{\log_{10} f_{\rm max}} d\log_{10} f P(\log_{10} f)  \label{eqn:pgw} \\ &\times& \int_{\log_{10} \epsilon{_{\rm UL}}}^0 d\log_{10}\epsilon P(\log_{10} \epsilon) \nonumber
\end{align}
where $P(\log_{10}f)$ and $P(\log_{10}\epsilon)$ are the probability density functions for \gw frequency and ellipticity, respectively, and $f$ has units of Hz. Also, we take $f_{\rm min}=120$ Hz and $f_{\rm max}=2000$ Hz, which is in the range of frequencies analyzed in the all-sky search \cite{LIGOScientific:2022lsr} with a cutoff at $f_{\rm rot}=60$ Hz to ensure we are targeting \msps.
% \yz{Here we treat GC and all-sky searches equally. Are the frequency ranges the same for both search?} \com{10-2000 Hz for GC and 10-2048 Hz for all-sky. So mostly yes}
% This corresponds to a \msp rotational frequency range of $f_{\rm rot}=[60,1000]$ Hz. 
% \yz{Why this range?} \com{Practically, the search covers from 20-2000 Hz in GW frequencies. } 
The distributions over ellipticity and frequency are normalized to one and assumed to be independent.

At last, the number of \msps detectable with \gws $N_{\rm GW}$ can be easily determined as $N_{\rm GW}=\pgw N_{\rm MSP}$,
where $N_{\rm MSP}$ is obtained in Sec. \ref{lumin-func}. If a set of luminosity function parameters $L_0$ and $\sigma_L$ leads to $N_{\rm GW}\geq 1$, it indicates that the Frequency-Hough all-sky search should have observed at least one \msp in the population if those $L_0$ and $\sigma_L$ did explain the GeV excess. Consequently, such a set should be excluded. 

% {In addition to the all-sky continuous \gw search, the number of \msps to explain the GeV excess may also be constrained by the results from the directed search towards the galactic center \cite{LIGOScientific:2022lsr}. In such a search, only \msps within one pixel are covered, which is not optimal; however, the directed search can use $\TFFT$ lengths about an order of magnitude higher than the all-sky ones, improving the sensitivity . }
% A correction factor to ``specialize'' the all-sky search results to the \gc is described in App. \ref{app:all-sky_ul}. We emphasize that a more careful study of how the upper limit on ellipticity would change as a function of the sky pixels around the \gc would be desirable, though the results would only differ by $\mathcal{O}(1\%)$.
% \yz{comment on the subtlety in directionality.} \com{what do you have in mind here? just that the UL will change slightly based on which pixel we look at, and that a proper search would look for \gwh emission following a certain pattern arising from the \gc depending on the number of neutron stars in a given pixel?} \yz{The discussion here singles out the GC search and did not comment on the conversion factor. Here we may want to comment on the pros and cons for each search, as well as the way to optimize the search in the future.}

% \input{table_pgw}

% \begin{equation}
% \int_{-\infty}^{+\infty }d\log_{10} f P(\log_{10} f)\int_{-\infty}^{+\infty } d\log_{10}\epsilon P(\log_{10} \epsilon)=1
% \end{equation}

% \section{All-sky search}

\section{Results}\label{sec:results}

\begin{figure*}[htb]
    \centering
    \includegraphics[width=\columnwidth]{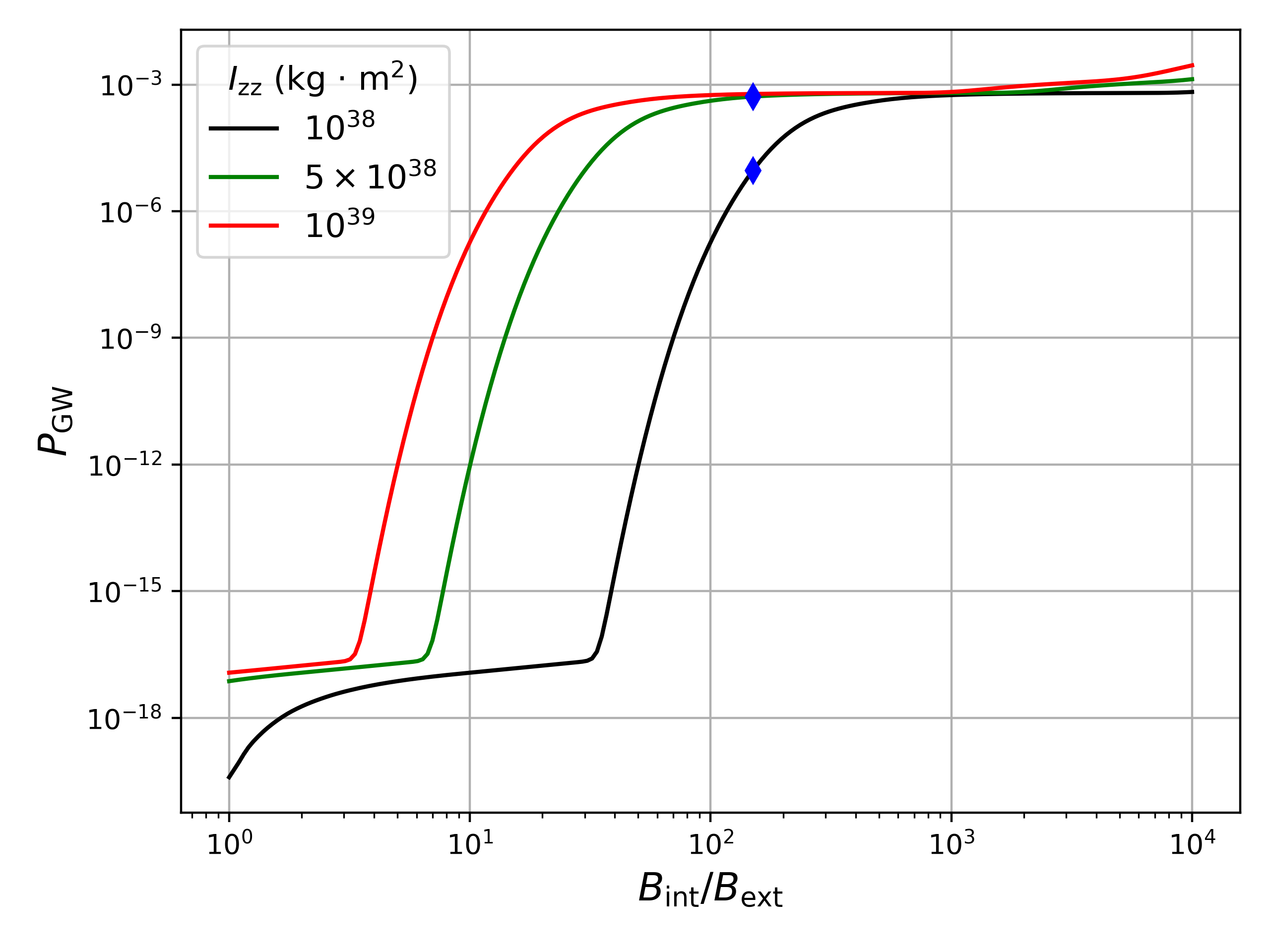}
    \includegraphics[width=\columnwidth]{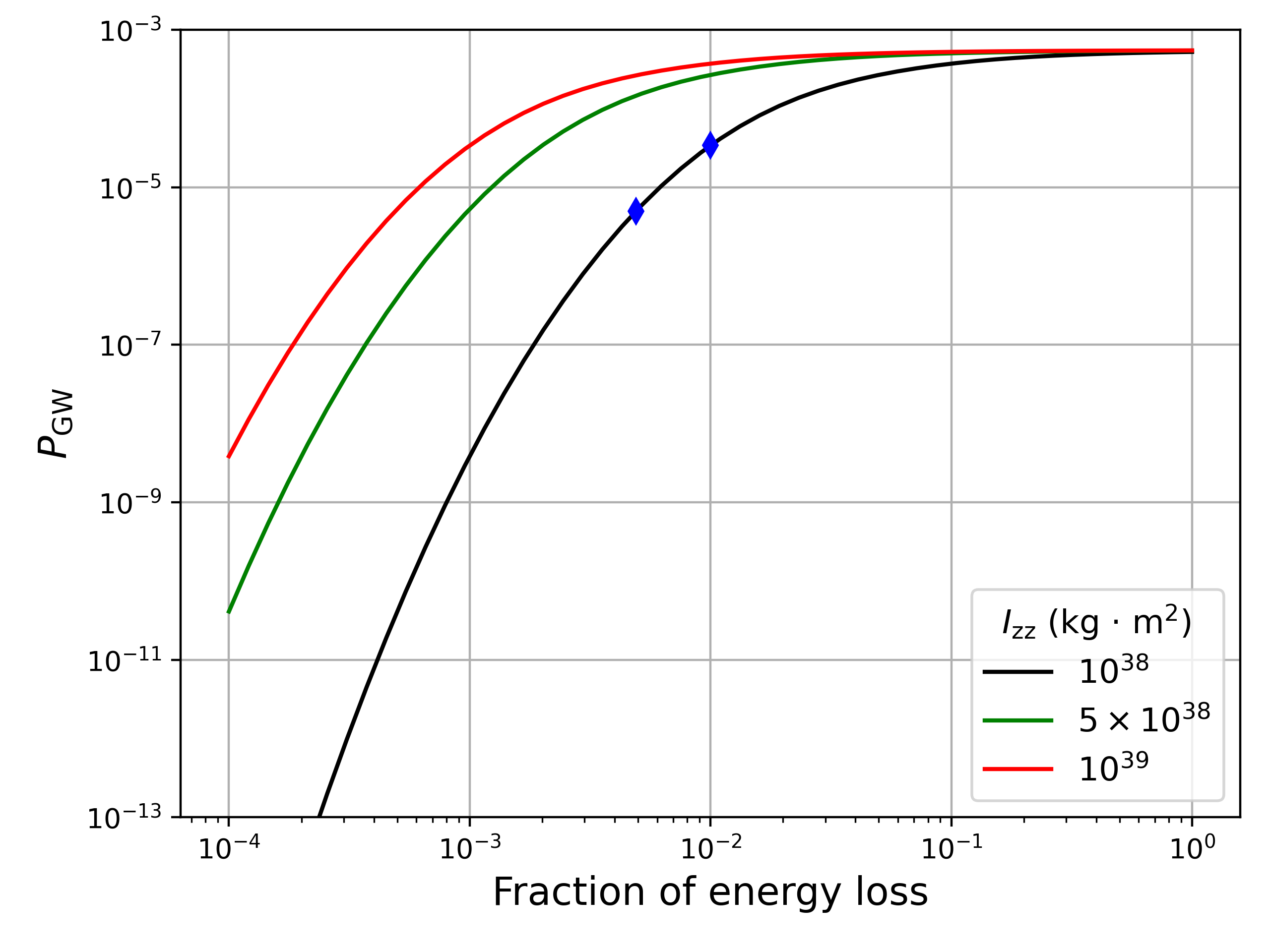}
    \caption{Probability to detect a \gwh signal obtained through Eq. (\ref{eqn:pgw}), as a function of the internal/external magnetic field ratio (left), and the fraction of rotational energy that we allow to be emitted as \gws (right). The blue diamonds denote the benchmarks that we used to perform concrete analyses with the chosen pulsar luminosity function, see Fig. \ref{fig:two_lum_func_sdlim_1perc}, \ref{fig:magnetic_strain_model} and \ref{fig:0.5perc}. $d=8$ kpc.}
    \label{fig:pgw}
\end{figure*}

$\pgw$ is fixed by the frequency and ellipticity distributions that we choose, as well as the upper limits on ellipticity from the Frequency-Hough all-sky search. It is therefore \emph{independent} of the luminosity function model considered to explain the GeV excess. Thus, we first present our results in terms of $\pgw$ which can be directly applied to probe the parameter space of \emph{any} luminosity function. Here, we provide, in the left- and right- hand side of Fig. \ref{fig:pgw}, $\pgw$ as a function of the ratio of internal and external magnetic fields, and of the percentage of rotational energy responsible for \gwh emission, respectively, for different choices of the moment of inertia. 
% Note that there does not exist a distribution over possible moments of inertia of the neutron stars, due to the large uncertainties in the equation of state; hence, we only consider fixed values here spanning an order of magnitude.
% \yz{This again is only for the GC search.} 

In the left panel of Fig. \ref{fig:pgw}, we see a smooth increase in $\pgw$ as the internal magnetic field strength grows, which corresponds to the peak in the first ellipticity PDF in App. \ref{app:dist}, shifting more and more to the right and thus allowing more support for higher ellipticites.  There is little difference beyond $\Bint/\Bext=10^3$ because at this value, the small ``bump'' in this PDF (at $10^{-8}$) already contributes to  $\pgw$, and the upper limits themselves are not sensitive enough to reach the peak ellipticity in the PDF. {Our results are very sensitive to the tail of the distribution of the ellipticity PDF, since the upper limits tend to only comprise the two small bumps at $10^{-8}$ or $10^{-6}$.}  
% \yz{What about the bump at $10^{-6}$? We may want to comment on the sensitivity to the tail distribution here.}  \com{done - what do you think?}
% \com{I addressed this point, so commenting it out}
Furthermore, at $\Bint/\Bext=10^4$, only a factor of $\sim 6$ separates the black and red curves, because even with the order of magnitude improvement in the upper limits when allowing $\Izz=10^{39}$ kg$\cdot$m$^2$  versus $\Izz=10^{38}$ kg$\cdot$m$^2$, the highest peak in the ellipticity PDF still does not contribute to $\pgw$. 
% \yz{I do not quite understand the comment here.} \com{I'm trying to say that the UL ellipticites are  still not low enough to include the largest peak in the ellipticity distribution. If that makes sense, I will try to rephrase..}
% \com{I think we talked about this, and I have slightly rephrased, so commented it out..}

In the right panel of Fig. \ref{fig:pgw}, when all rotational energy goes into \gws, the moment of inertia does not play any role in determining $\pgw$. An order of magnitude increase in the moment of inertia only increases the possible ellipticity by a factor of $\sqrt{10}\sim3$, which does not alter much the second ellipticity PDF given in App. \ref{app:dist}. Furthermore, allowing less and less rotational energy to be converted into \gws has a sizable effect, since the ellipticity PDF in Fig. \ref{fig:pdf_crust} scales linearly with this fraction. Again, due to the sensitivity of $\pgw$ to the tail of the ellipticity PDF, $\Izz$ matters more for lower fractions than for higher ones. 
% \yz{I get myself confused. Shouldn't this result be independent on $I_{zz}$? After we fix the fraction, it directly gives us the GW amplitude, which is the quantity we measure. Why does it care about $I_{zz}$?} \com{the fraction doesn't depend on $\Izz$, but the upper limits do. So, in equation \ref{eqn:pgw}, the bound on the integral changes for different $\Izz$}
% \com{I think we talked about this, so commented it out..}

\begin{figure*}[htb]
    \centering
     \includegraphics[width=\columnwidth]{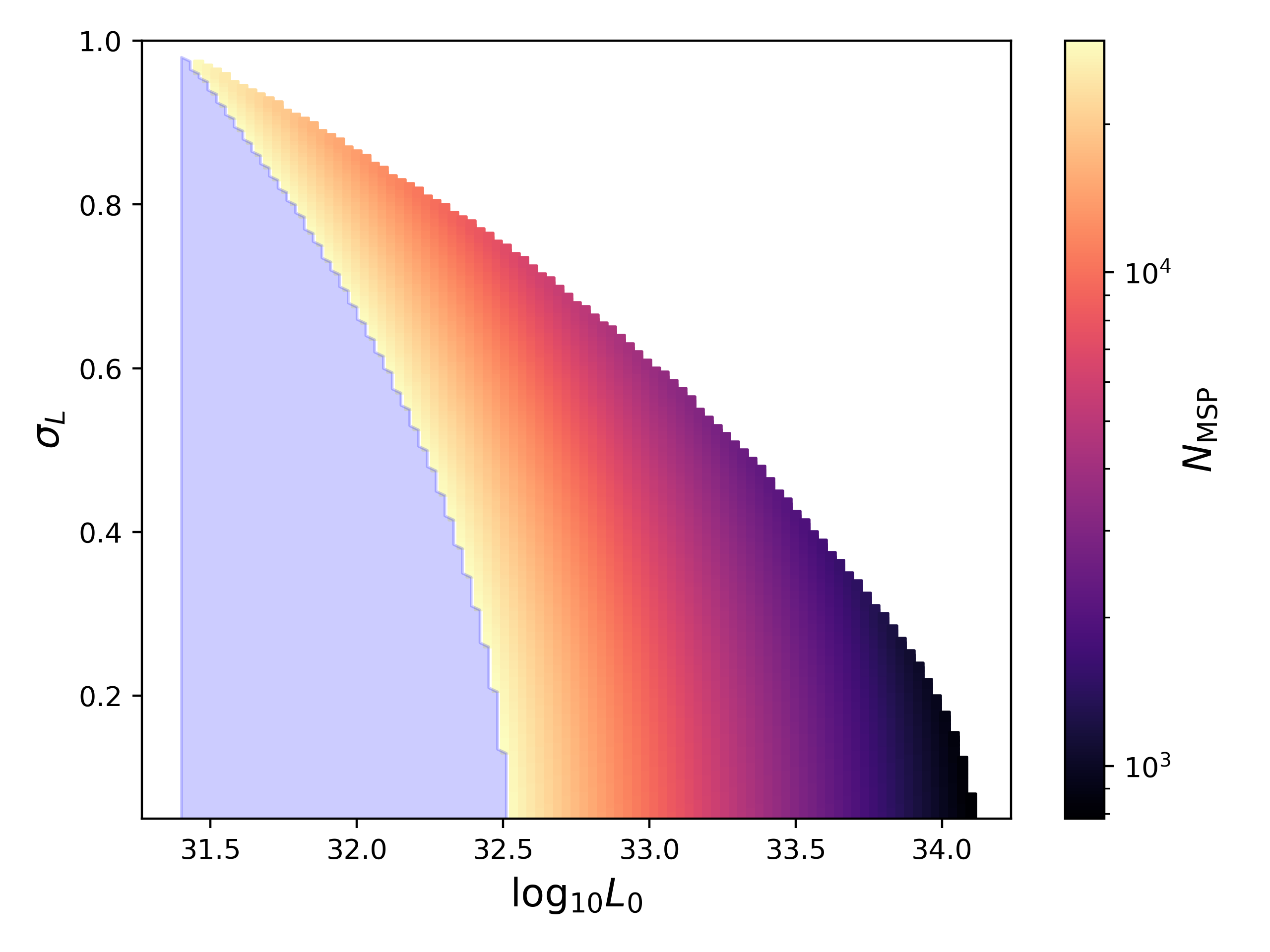}
    \includegraphics[width=\columnwidth]{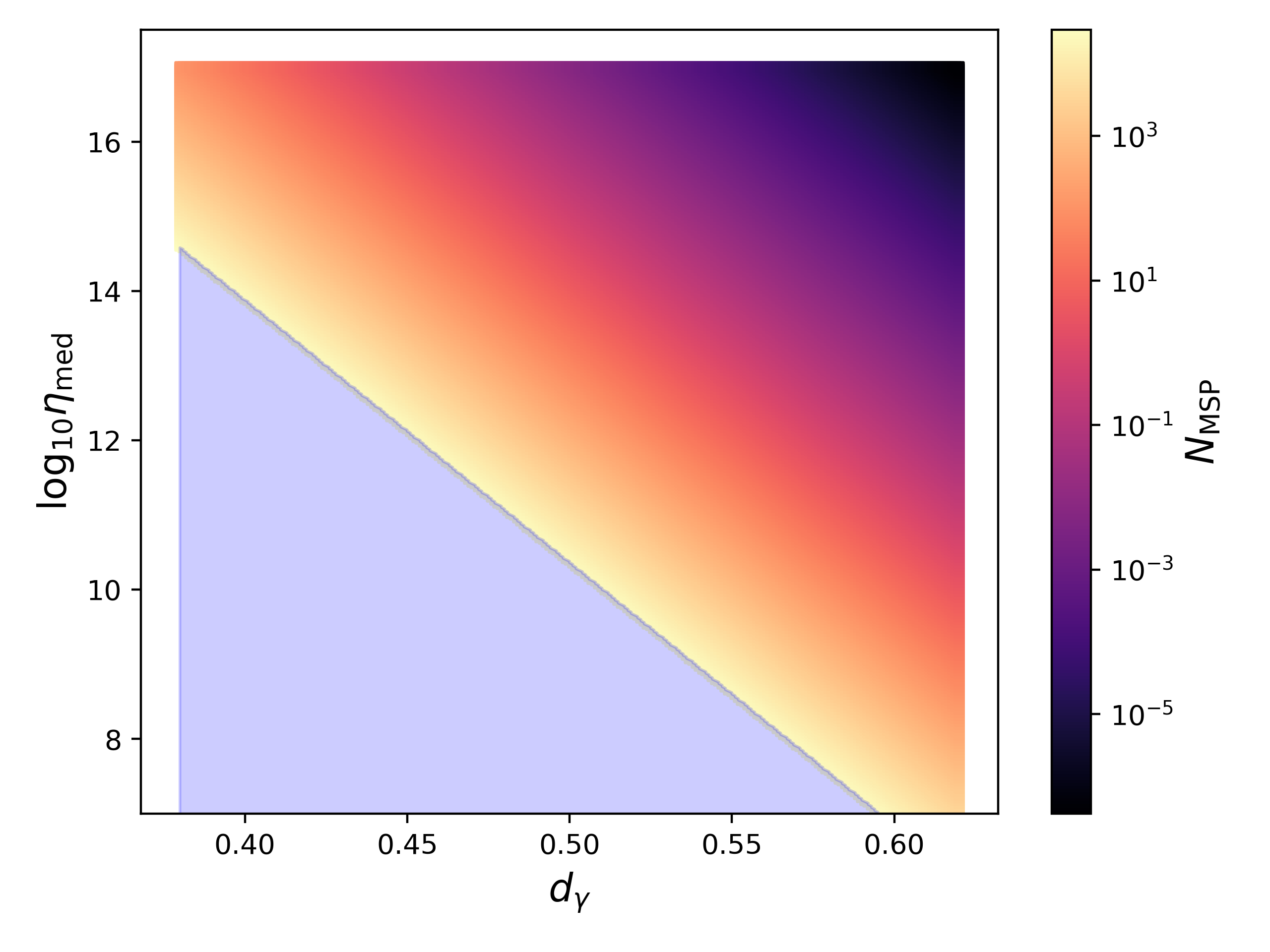}
    \caption{Exclusion regions (light blue) based on the upper limits of the O3 Freuquency-Hough all-sky search for the log-normal (left) and power-law (right) luminosity functions, employing a probability density function for the ellipticity that assumes 1\% of the rotational energy loss of the star goes into \gws. We take $d=8$ kpc and $I_{\rm zz}=10^{38}$kg$\cdot$m$^2$ in this analysis.  
    % The boundary of the parameter space is obtained by requiring (1) the \msps above the gamma-ray detection threshold to be fewer than 20\% of all the 4FGL-DR2 sources; and (2) the ratio of the flux emitted by the resolved point sources to be smaller than 20\% of the total flux from the \gc. 
    The upper-right white region on the left-hand plot has been excluded by \fermi.
    }
    \label{fig:two_lum_func_sdlim_1perc}
\end{figure*}

As an example of what one can do with specific values of $\pgw$, we present our results in terms of exclusion regions in the parameter space of a pulsar luminosity function that explains the GeV excess given in Eq. (\ref{eqn:log-normal}) and Eq. (\ref{eqn:power-law}), in the left- and right-hand panels of Fig. \ref{fig:two_lum_func_sdlim_1perc}, respectively. Here, we consider as a benchmark the ``fixed fraction of \gwh energy loss'' model, in which 1\% of the star's rotational energy is converted into \gws, given by a blue diamond in Fig. \ref{fig:pgw}. If a pair of luminosity function parameters (e.g. $L_0$,$\sigma_L$ in Eq. (\ref{eqn:log-normal}), left panel, and $\eta_{\rm med},d_\gamma$ in Eq. \ref{eqn:power-law}, right panel) results in to too many \msps in the \gc, such that at least one of them would have been detected in the O3 Frequency-Hough all-sky search, we can rule out that point in the luminosity function parameter space. The benchmarks of a chosen set of parameters are labeled as blue diamonds in this figure, and are more explored in the App. \ref{app:vary}, along with exclusion regions showing the permutations of $a_\gamma,b_\gamma,$ and $d_\gamma$.

\section{Conclusions}

In this work, we have, for the first time, presented \gwh constraints on the \msp hypothesis to explain the observed GeV excess by \fermi using LIGO/Virgo/KAGRA data. We used upper limits from the most recent Frequency-Hough all-sky search to calculate the number of detectable \msps for various luminosity function parameters, integrating over physically-motivated distributions of rotational frequency and ellipticity. We considered two models for neutron-star deformation, one in which an internal magnetic field sustained the deformation, and another in which we were agnostic to the mechanism of deformation and instead assumed a fixed fraction of rotational energy loss via \gws. {For these two models, we computed the probability of detecting a \gwh signal in a search of O3 data, and then excluded portions of the luminosity function parameter space for both models on the basis of null observations for particular parameter choices.}

{Our exclusion regions depend on 
(1) the ellipticity distributions we calculate, which themselves are functions of the external magnetic field or the fraction of the rotational energy we assume goes into \gws, (2) the \gwh frequency distribution we use, and (3) the moment of inertia. } Furthermore, our results are valid for isolated neutron stars, which comprise about half of all known pulsars (the other half are in binary systems). We show, however, in App. \ref{app:mspbin}, that our results would apply to \msps in binary systems with certain orbital parameters, given in Fig. \ref{fig:binary_parm_space}.

In the future, all-sky \gwh searches will be able to dig deeper into the noise at all frequencies, thus our constraints could be greatly improved. 
% \yz{The \msp distribution dies off at higher frequency, I am not sure this is a right statement.} \com{That is true, but at lower frequencies, the number of detectable pulsars is already less than 0, before integrating over the frequency distribution. It's shown in those ``intermediate'' plot}
A factor of $5-10$ improvement in the upper limits, expected in third-generation detectors \cite{punturo2010einstein,reitze2019cosmic}, would allow us to almost completely exclude this luminosity function under the assumptions presented in this work. {Furthermore, we plan to devise a template-based method to search for the GeV excess in the \gc by weighting sky pixels by the expected spatial distribution of \msps in each one. This approach should improve the sensitivity to \msps, and would allow us to incorporate other aspects of \msp astrophysics into our work, such as frequency/ellipticity distributions as a function of location.  Our future work would also require using finer-resolution sky pixels to explore the inner parsec regions of the \gc, which would add to the computational cost of such \gwh searches and thus needs to be studied. }
% \yz{Can we comment more on other aspects of future improvements?} \com{added some additional text. What do you think?}
% \com{we discussed this I think}

We also plan to combine the constraints inferred by Fermi and \gwh detectors to probe more portions of the luminosity function parameter space. Furthermore, we could employ correlated frequency/ellipticity probability density functions, e.g. \cite{Haskell:2022pqt}, 
% \yz{are there refs for this correlation already?} \com{I don't think so - but actually this was just posted on the DCC today: https://dcc.ligo.org/P2200310, see equation 27, where $\epsilon$ depends on $\sqrt{B}$, instead of just $B$. We could incorporate this result too} 
that should be more representative of the underlying physics of \gwh emission from \msps. 
% Finally, we could use our results to potentially rule out various neutron star equations of states by varying the moment of inertia, or using astrophysical luminosity functions. 
% \yz{I do not see the relation to our study in this paper.} \com{I think it's a weak connection.. I have removed it } 
Our work therefore represents a major connection between neutron stars, \gws, dark matter, and the \gce.

\section*{Acknowledgments}

This material is based upon work supported by NSF's
LIGO Laboratory which is a major facility fully funded
by the National Science Foundation.

We thank Bryn Haskell, Ian Jones and Graham Woan for helpful discussions regarding how to construct ellipticity distributions from known \msps. We also acknowledge the \lvk continuous-wave group for helpful feedback on this project, including Rodrigo Tenorio for questions regarding the application of these results to \msps in binary systems.

All plots were made with the Python tools Matplotlib \cite{Hunter:2007ouj}, Numpy \cite{Harris:2020xlr}, and Pandas \cite{mckinney-proc-scipy-2010,reback2020pandas}.

A.L.M. is a beneficiary of a FSR Incoming Post-doctoral Fellowship. Y.Z. is supported by the U.S. Department of Energy under Award No. DESC0009959.

% F.D.L. is supported by a FRIA grant from the Fonds de la Recherche Scientifique FNRS, Belgium. 

We would like to thank all of the essential workers who put their health at risk during the COVID-19 pandemic, without whom we would not have been able to complete this work.

\appendix 

% \section{Upper limits from \gc search}

% We provide in this section the upper limits we employ in this analysis from the \gch search in O3 \cite{LIGOScientific:2022lsr}.

% \begin{figure}
%     \centering
%     \includegraphics[width=0.49\textwidth]{figures/O3_GC_ellip_UL.png}
%     \caption{Upper limits on ellipticity as a function of \gwh frequency from a search towards the \gc in the third observing run of LIGO/Virgo/KAGRA. $d=8$ kpc, $\Izz=10^{38}$ kg$\cdot$m$^2$}
%     \label{fig:ellip_UL_O3}
% \end{figure}

\section{Frequency and Ellipticity distributions}\label{app:dist}
\begin{figure}
    \centering
    \includegraphics[width=0.49\textwidth]{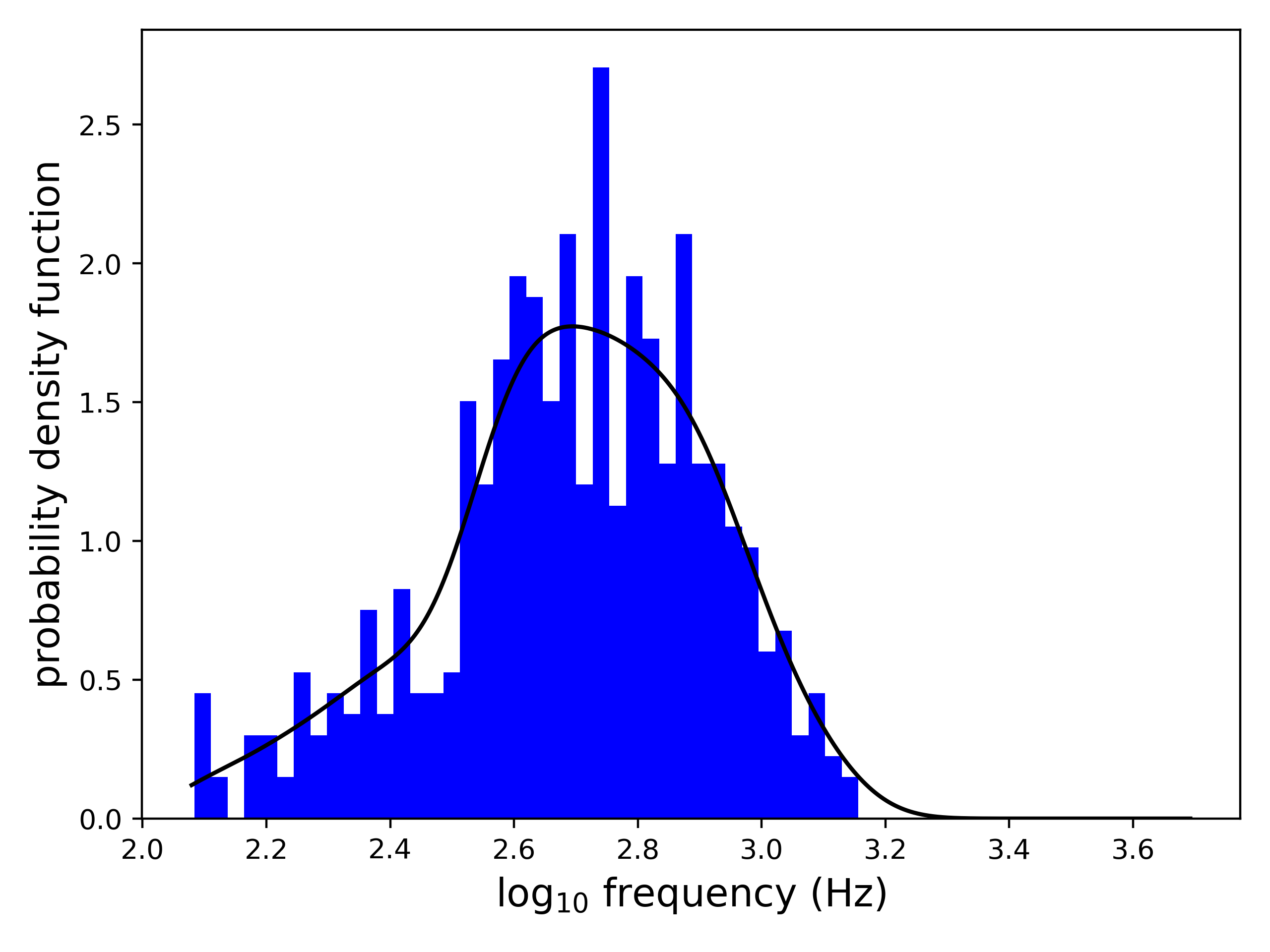}
    \caption{Gravitational-wave frequency probability density function, with a minimum frequency cutoff at $f_{\rm rot}=60$ Hz, ($f_{\rm GW}=120$ Hz) derived from the ATNF catalog \cite{Manchester:2004bp}. The black line is a Kernel density estimator fit to the histogram \cite{kde}. }
    \label{fig:f_pdf}
\end{figure}

\begin{figure}
    \centering
    \includegraphics[width=0.49\textwidth]{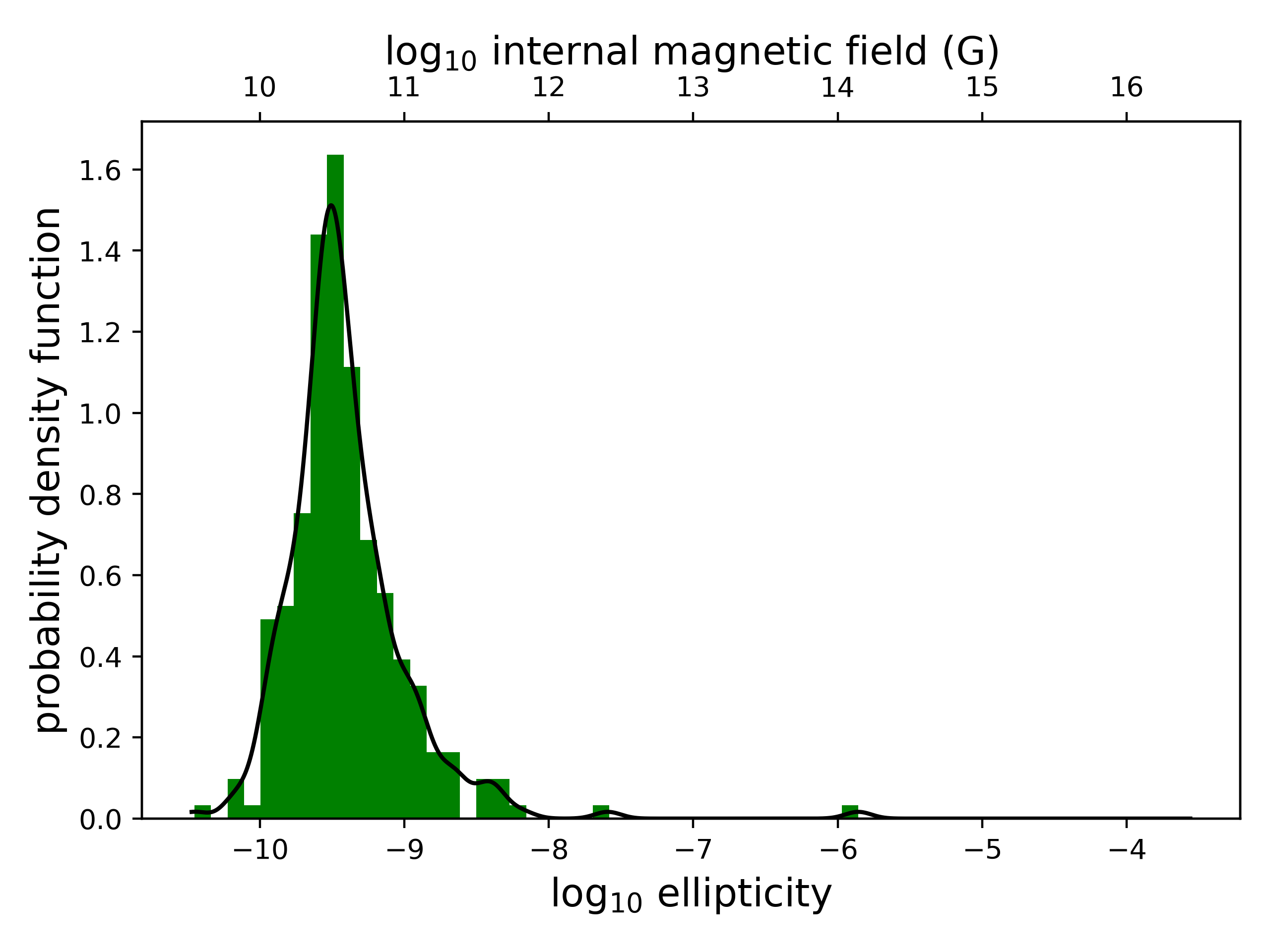}
    \caption{Here we show the ellipticity probability distribution, assuming the deformation of a \msp caused by an internal magnetic field. We take the probability distribution of the external magnetic field to be the same as that in the ATNF catalog \cite{Manchester:2004bp}, and we use the relation $\Bint=150\Bext$. The black line is a Kernel density estimator fit to the histogram \cite{kde}. {For reference, the best ellipticity upper limits on some known \msps, e.g. J0437--4715, J0711--6830 and J0737--3039A, are $8.3\times 10^{-9}, 7.2\times 10^{-9}$, and $1.0\times 10^{-6}$ \cite{LIGOScientific:2020gml}.} }
    \label{fig:ellip_pdf_I_1e38_factor_100}
\end{figure}

\begin{figure}
    \centering
    \includegraphics[width=0.49\textwidth]{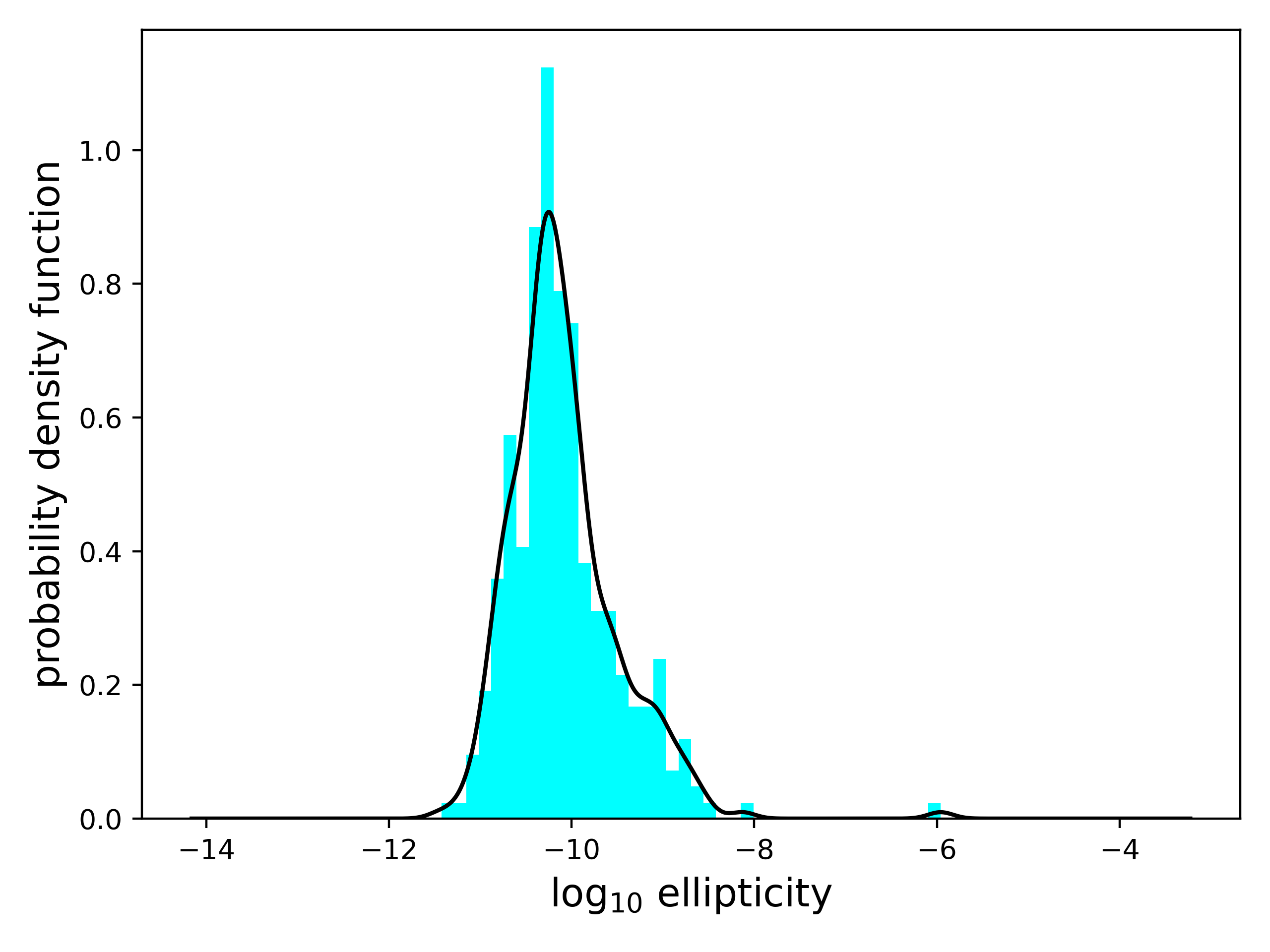}%.png}
    \caption{Ellipticity probability density function obtained when assuming that \gwh emission accounts for 1\% of the star's rotational energy loss. $\Izz=10^{38}$ kg$\cdot$m$^2$. The black line is a Kernel density estimator fit to the histogram \cite{kde}.}   
    \label{fig:pdf_crust}
\end{figure}

Here we present the frequency and ellipticity distributions derived from the ATNF catalog. The frequency distribution for \msps is given in Fig. \ref{fig:f_pdf}, with a required minimum value of $f_{\rm rot}=60$ Hz. The ellipicity distributions for each deformation model considered -- magnetic strain and fixed fraction -- are given in Figs. \ref{fig:ellip_pdf_I_1e38_factor_100} and \ref{fig:pdf_crust}, respectively. These distributions are used as inputs, along with the upper limits from an O3 all-sky search, to determine the number of detectable \msps in O3.
{If we were to use a different frequency distribution as given in \cite{Ploeg:2020jeh}, we do not expect the exclusion regions to change much, since the integration over the frequency distribution given in Eq. \ref{eqn:pgw} always results in a constant that varies by $\mathcal{O}(1)$. When using the distribution in Fig. \ref{fig:f_pdf}, we obtain $\int_{\log_{10} f_{\rm min}}^{\log_{10} f_{\rm max}} d\log_{10} f P(\log_{10} f)\sim 0.8$ and for that given in \cite{Ploeg:2020jeh}, $\sim 1$. }

In the magnetic case, we note that larger $\Bint$ would correspond to bigger ellipticities, which would improve the exclusion regions presented in this work, so even though Fig. \ref{fig:ellip_pdf_I_1e38_factor_100} quotes a value of $\Bint=150\Bext$, this is a relatively conservative choice with respect to the range of possible ratios (at most $10^4$).

For a self-consistency check, \gws would only comprise approximately $\mathcal{O}(0.1\%)$ of the rotational energy of the star, which is calculated using the known $\dot{f}_{\rm rot}$ parameters from the ATNF catalog and comparing that to the \gwh induced spin-down.

\section{Luminosity Function details}
\label{app:lum}

We give more details here about the luminosity functions. The first benchmark luminosity function in Eq. (\ref{eqn:log-normal}) is relatively straightforward. There are two free parameters $L_0$ and $\sigma_L$.  As a comparison with the results in \cite{Dinsmore:2021nip}, we further require the following two criteria to be satisfied. First, given the number of \msps needed to explain the GeV excess, one should not have too many of them that are above the sensitivity threshold for \fermi 4FGL-DR2. Here we require this to be fewer than 20\% of all the 4FGL-DR2 sources, i.e. the number of detectable \msps is smaller than 53. Second, we require that the flux of the gamma ray excess at the \gc is not reduced significantly by masking the \fermi 4FGL-DR2 point sources \cite{DiMauro:2021raz}. This demonstrates that the GeV excess should not be dominantly coming from these identified point sources.%, and motivates us to impose a 20\% cut on $N_r$ and $R_r$. 
Thus we require the ratio of the flux emitted by the resolved point sources to be smaller than 20\% of the total flux from the \gc.

The second benchmark in Eq. (\ref{eqn:power-law}) is more general, but it is also quite involved, so let us provide some details.
Here $E_{\rm cut}$, $\eta$ and $B$ are assumed to follow the log-normal distribution. The energy emission rate, $\dot E$, can be written as $\dot E = 4\pi^2 I_{zz} \dot P/P^3$, where $P$ is the period of a \msp. After averaging on the angle between the rotation and magnetic field axes$, \dot P$ can further be written as $\dot P = 5\pi^2 R_{\rm MSP}^6 B^2/(3c^3 I_{zz}P)$. The median of $E_{\rm cut}$ is related to $\dot E$ as 
\begin{eqnarray}
    \log_{10}\left(\frac{E_{\rm cut,med}}{{\rm MeV}}\right)= 
    a_{E_{\rm cut}}\log_{10}\left(\frac{\dot E}{10^{34.5} \rm erg\ \rm s^{-1}}\right)+b_{E_{\rm cut}}. \nonumber
\end{eqnarray}
In \cite{Ploeg:2020jeh}, all undetermined parameters are obtained by fitting to the GeV excess, see their Table 4 for more details. In our study, we choose their Model A1. Except for $\eta_{\rm med}$ and the power indices \{$a_{\gamma}$, $b_{\gamma}$, $d_{\gamma}$\}, we take the central values for the rest of the parameters in the luminosity function. We present our results in each pairs of $\eta_{\rm med}$ v.s. one power index, while fixing the other two as their central values.

Following the discussion in \cite{Dinsmore:2021nip}, and assuming the luminosity function of \msps in the \gc does not have any spatial dependence, the conversion from the total luminosity to the total gamma-ray flux is 
\begin{align}
    \frac{F_\text{GCE}}{L_\text{GCE}} &= \frac{1}{4\pi}\left[\int_{\Omega}d\Omega\int_0^\infty ds \rho_\text{GCE} (r)\right] \times \\
    & \ \ \left[\int_{\Omega}d\Omega \int_0^\infty s^2 ds \rho_\text{GCE} (r)\right]^{-1} \nonumber \\
    &= \SI{1.11e-46}{\per\centi\meter\squared}.
    \label{eqn:f-to-l}
\end{align}
Here, $s$ and $r$ are the distances from the Earth and \gc to the point of integration, respectively, and $\rho_{\rm GCE}$ is the number density distribution of the \msps. We take $\rho_{\rm GCE}$ to be the generalized NFW profile, which is determined by fitting the GeV gamma-ray excess in the \gc,
\begin{align}
    \sqrt{\rho_{\text{GCE}}(r)}\propto (r/r_s)^{-\gamma}(1+r/r_s)^{-3+\gamma}.
\end{align}
We choose $\gamma=1.2$ and $r_s=20$ kpc \cite{DiMauro:2021raz,Calore:2014xka,Gordon:2013vta}.  We note that this conversion factor is not very sensitive to the \msp spatial distribution: it remains almost the same even if all \msps are concentrated at the \gc.

The value for the total flux of the \gce can be obtained from \textit{Fermi's} measurement. However, the details during the data analysis, such as the fitting procedure and background modeling etc, may lead to a difference of $\mathcal{O}(1)$. A detailed comparison can be found in \cite{Dinsmore:2021nip}. In this study, we make a generic choice and set $F= 10^{-9}$ erg/cm$^2$/s. Using Eq. (\ref{eqn:f-to-l}), we get $L_\text{GCE}\approx 10^{37}$ erg/s.
Consequently, by integrating over the function, we obtain values of $N_{\rm MSP}$ as a function of various model parameters. This gives us the required number of \msps in order to explain the GeV excess observed at the \gc.  
Later, $N_{\rm MSP}$ will be tested based on the results from the Frequency-Hough all-sky search for continuous waves in O3, see Sec. \ref{sec:meth}.

% Later, $N_{\rm MSP}$ will be tested based on the continuous \gwh results of the Frequency-Hough all-sky search in O3, see section \ref{sec:meth}.

\section{Upper limits from Frequency-Hough all-sky search and the conversion factor to a directed search}\label{app:all-sky_ul}

\begin{figure}
    \centering
    \includegraphics[width=0.49\textwidth]{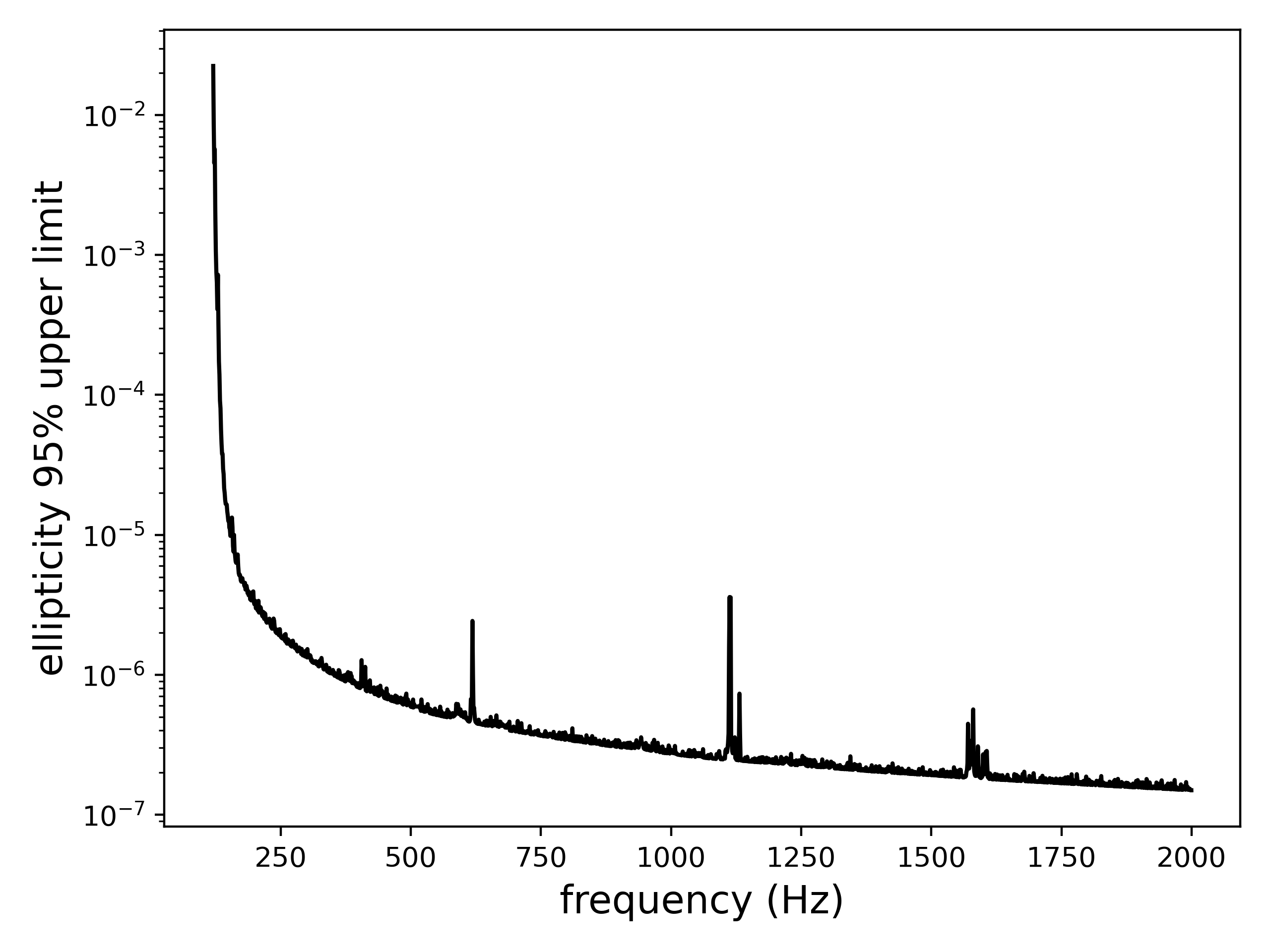}
    \caption{Upper limits on ellipticity as a function of \gwh frequency from the Frequency-Hough all-sky search in the third observing run of LIGO/Virgo/KAGRA. One sets $d=8$ kpc, $\Izz=10^{38}$ kg$\cdot$m$^2$, to generate these limits.}
    \label{fig:ellip_UL_O3}
\end{figure}

% \yz{I think I understand how the conversion factor is calculated now. Thanks! I may ask more questions later.} 
We provide here the Frequency-Hough all-sky upper limits on ellipticity from \cite{LIGOScientific:2022pjk}, in Fig. \ref{fig:ellip_UL_O3}, averaged over sky location. These limits, initially derived in terms of strain as a function of frequency, are translated to limits on the ellipticity using Eq. \eqref{eq:h0}.
% The \gch search targets a specific pixel in the sky, with a size of $\mathcal{O}(10-100$) pc, depending on the frequency. It is therefore worth asking how our results would change if we consider the more realistic case in which \msps exist across the \gc, i.e. across many sky pixels. 
% \yz{We may want to comment on this subtlety earlier.} \com{agreed.} 
% \yz{I still have a hard time to formulate the conversion factor. Can we provide some equations for illustration here?} \com{how about what is written below?}

Furthermore, in order to calculate the conversation factor needed to ``specialize'' the sky-averaged limits to the pixels surrounding the \gc, we start with the following (abridged) formula used in \cite{LIGOScientific:2022pjk} that relates the geometric factor $\mathcal{S}$ to how a monochromatic \gw couples to the detector:

\begin{equation}
\mathcal{S}^2=\left(A_+F_+ + A_\times F_\times\right)^2.
\end{equation}
{$F_+$ and $F_\times$ are the time-dependent detector beam pattern functions that convey directional and temporal sensitivity to all sky locations, which are:}
\begin{equation}
\begin{aligned}
F_+(t)=a(t)\cos 2\psi + b(t)\sin 2\psi\\
F_\times(t)=b(t)\cos 2\psi - a(t)\sin 2\psi.
\end{aligned}
\end{equation}
{$t$ is time, $a(t)$ and $~b(t)$ can be found in \cite{Jaranowski:1998qm} and depend on the detector and source location, and $\psi$ is the \gwh polarization angle. The terms $A_+$ and $~A_\times$ are:}
\begin{equation}
\begin{aligned}
A_+=\frac{1+\cos^2 \iota}{2}\\
A_\times=\cos \iota,
\end{aligned}
\end{equation}
% \yz{How to obtain this skymap? Also, is the choice of 2000 Hz conservative?} \com{It's described in here \cite{Astone:2014esa}. Conservative in what sense? 2000 Hz just gives the greatest pixelization of the sky, since the number of sky points scales with the square of the frequency. } 
{where $\iota$ is the inclination angle between the Earth and the neutron star. In \cite{LIGOScientific:2022pjk}, the upper limits are computed by taking the average over all the source parameters, which gives an overall factor of:}
\begin{equation}
\mathcal{S}^2_{\alpha,~\delta,~\psi,~\iota}=<\mathcal{F}^2>_{\alpha,\delta,\psi,\iota}\simeq \frac{4}{25},
\label{eq:Fave}
\end{equation}
{approximately independent of which detector that is considered. $\alpha$, $\delta$ and $\psi$ are the right ascension, and declination, and polarization angle, respectively, of a source.} 

{Here, we only need to average over $\psi$, $\cos \iota$, and $t$ to obtain a sky-dependent factor:}

\begin{equation}
    \mathcal{S}^2_{~\psi,~\iota}=<F^2_+>_{t,\psi}<A^2_+>_{\iota}+<F^2_\times>_{t,\psi}<A^2_\times>_{\iota}
    \label{eq:Fave_t_psi_iota}
\end{equation}
{When we take the ratio of Eq. (\ref{eq:Fave}) to Eq. (\ref{eq:Fave_t_psi_iota}), we obtain a factor by which we multiply the upper limit in \cite{LIGOScientific:2022pjk} at a fixed frequency, to obtain a sky-dependent upper limit:} 
\begin{equation}
    \mathcal{C}(\alpha,\delta) = \sqrt{\frac{\mathcal{S}^2_{\alpha,~\delta,~\psi,~\iota}}{\mathcal{S}^2_{~\psi,~\iota}}}.
\end{equation}

\begin{figure}
    \centering
    \includegraphics[width=\columnwidth]{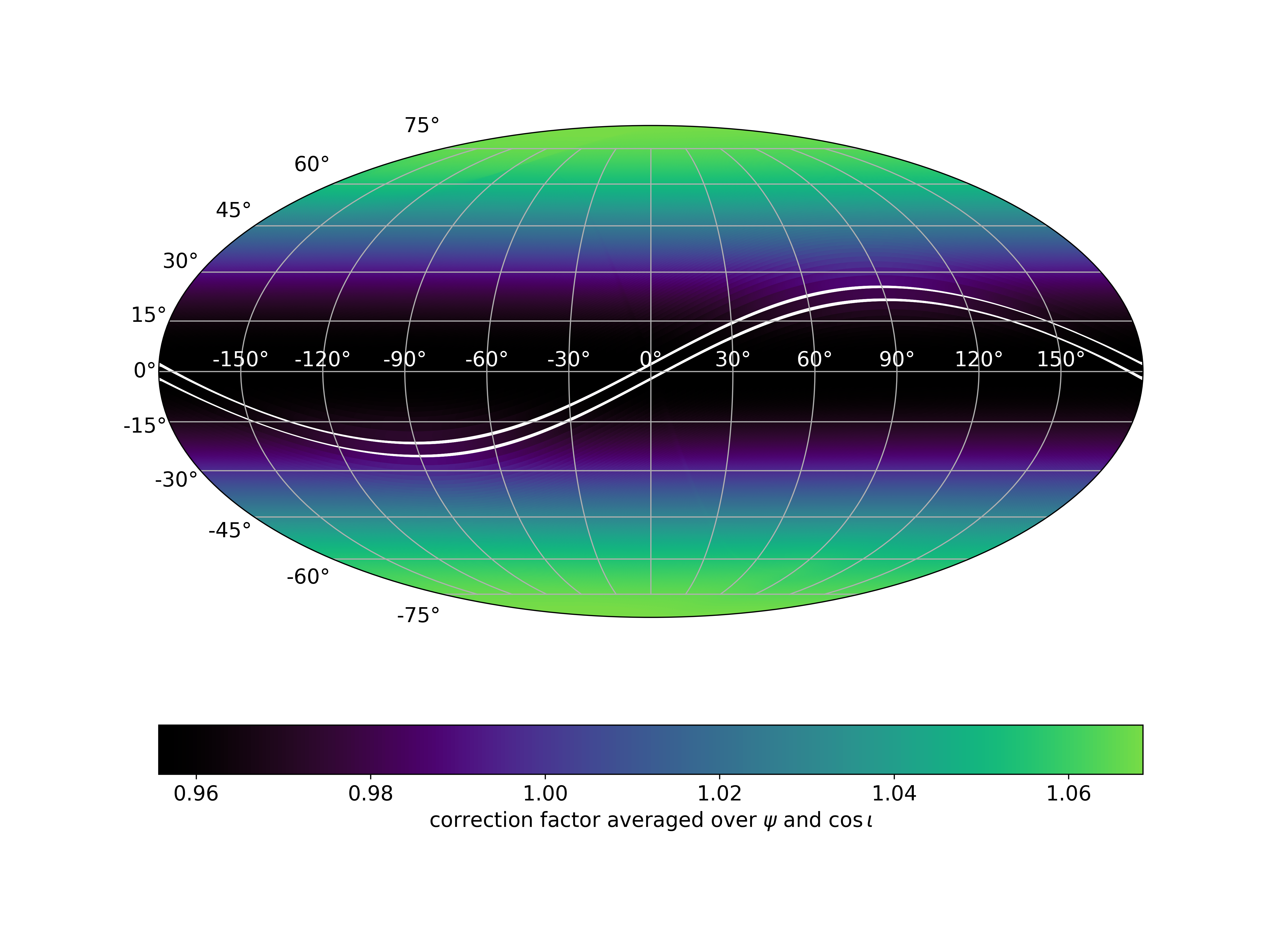}
    \caption{Here we show the factor by which upper limits from the Frequency-Hough all-sky search need to be multiplied to obtain sky-dependent ellipticity upper limits at 2000 Hz. Note that the sky grid is a function of frequency \cite{Astone:2014esa}.}
    % The upper plot is calculated for $\epsilon_s = 10^{-18}$ and the lower one for $\epsilon_s = 10^{-22}$.}
    \label{fig:skymap}
\end{figure}

In Fig. \ref{fig:skymap}, we produce a skymap at $f=2000$ Hz that shows $\mathcal{C}(\alpha,\delta)$. 
We can see that, for different pixels, $\mathcal{C}(\alpha,\delta)$ changes by no more than a few percent in either direction from 1, indicating that the upper limits used in one pixel are reflective of considering a larger number of sky pixels in the \gc. 
{Note that the skymap looks different at each frequency: the number of sky points scales with the square of the \gwh frequency, since the criterium to grid the sky requires that when moving from one point to another, the modulation induced by the Doppler effect is confined to one frequency bin \cite{Astone:2014esa}. } However, the percent change in the upper limits as a function of sky location remains about the same, regardless of the frequency.
% -- since sky resolution scales with the square of the frequency, since the criterium to bin the sky requires that when moving from one point to another, the modulation induced by the Doppler effect is confined to one frequency bin \cite{Astone:2014esa} \com{I have explaind more} -- the percent change in the upper limits as a function of sky location remains about the same.

\section{Millisecond pulsars in binary systems}
\label{app:mspbin}
\begin{figure}
    \centering
    \includegraphics[width=\columnwidth]{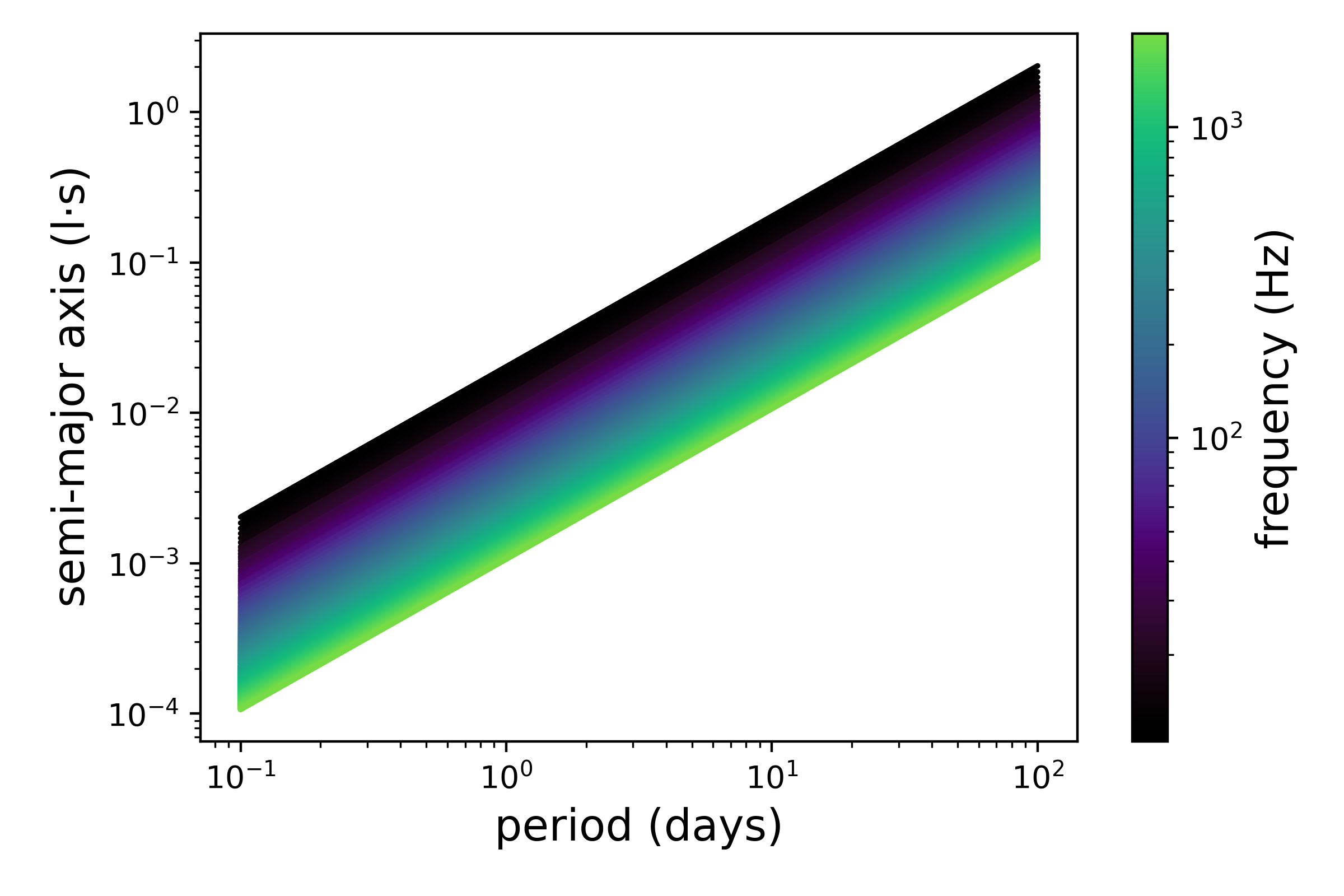}
    \caption{Parameter space of \msps in binary systems to which our constraints would apply. Note that only a small fraction of the orbital parameters of known \msps in binary systems lie within this range. }
    % The upper plot is calculated for $\epsilon_s = 10^{-18}$ and the lower one for $\epsilon_s = 10^{-22}$.}
    \label{fig:binary_parm_space}
\end{figure}

The results presented here are valid if \msps are isolated, but one expects that $O(1)$ fraction of them exist in binary systems \cite{Tauris:2017omb}. 
% We could use the upper limits from a search for unknown neutron stars in binary systems in O3 \cite{LIGOScientific:2020qhb} to calculate these exclusion regions, but only the frequency range 50-300 Hz is covered, while our constraining power comes from much higher frequencies. 
We thus compute the orbital parameters of binary systems to which our results could apply by requiring that the Doppler modulation induced by the orbital motion of the system is contained within one frequency bin, and neglecting orbital eccentricity \cite{LIGOScientific:2019yhl}:

\begin{equation}
a_p \leq 0.076\left(\frac{P}{1\text{ day}}\right)\left(\frac{f}{100 \text{ Hz}}\right)^{-1} \left(\frac{\TFFT}{1800 \text{ s}}\right)^{-1} \text { l$\cdot$s}.
\end{equation}
Here, $a_p$ is the semi-major axis with units light-seconds l$\cdot$s , $P$ is the orbital period, and $\TFFT$ is the Fast Fourier Transform length used in the search, which is also a function of frequency. Fig. \ref{fig:binary_parm_space} shows the orbital parameter space $a_p$ and $P$ to which we are sensitive, as a function of \gwh frequency. Our results are thus valid for \msps in binary systems whose orbital parameters lie within these ranges. We note, however, that about half of known \msps are found in binary systems, and of those, only a small fraction of known \msps lie within this parameter space.

\section{How exclusion regions vary based on $\Izz$, $d$ and fraction of energy loss }\label{app:vary}

\begin{figure*}[htb]
    \centering
    \includegraphics[width=\columnwidth]{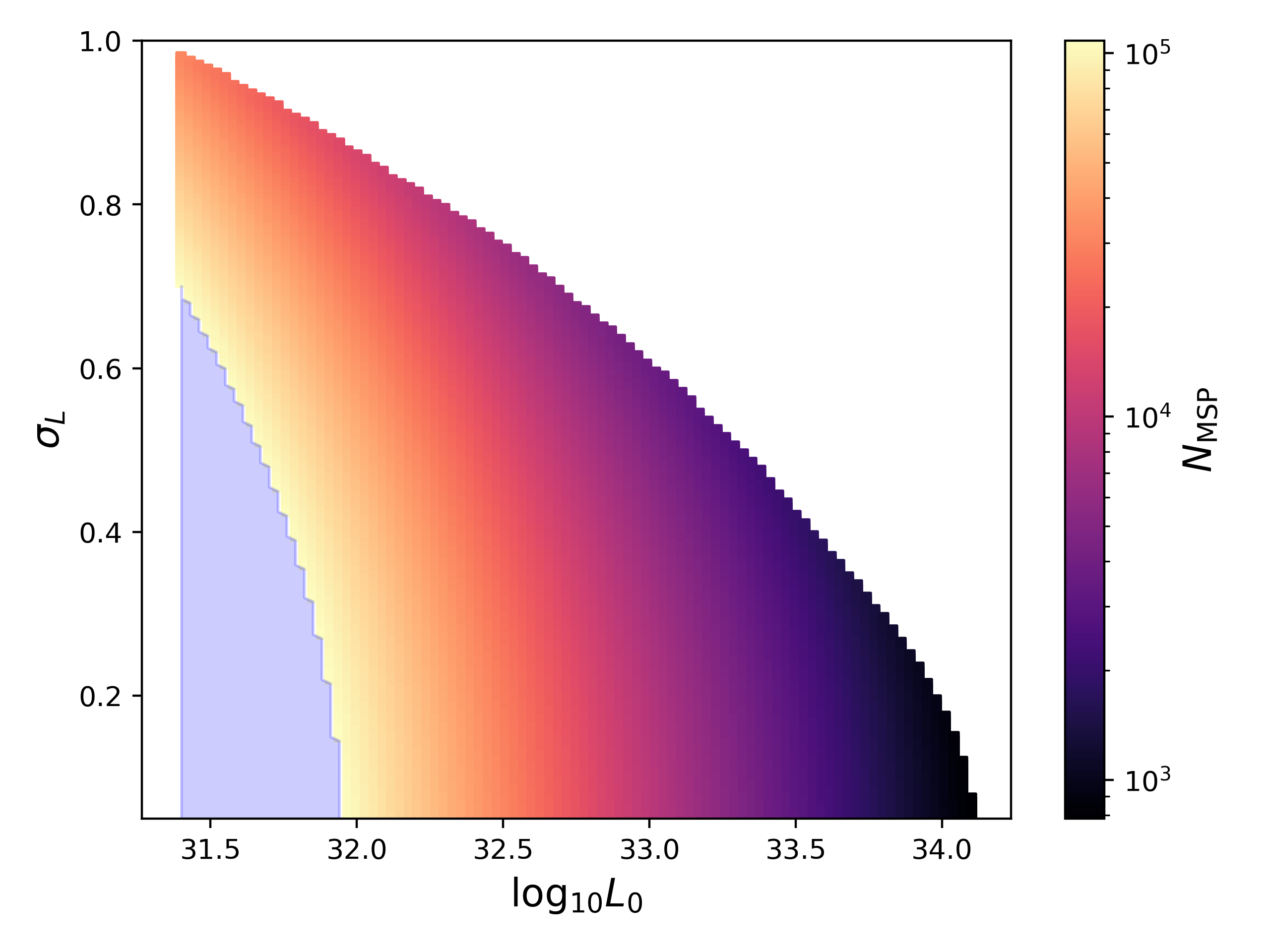}
    \includegraphics[width=\columnwidth]{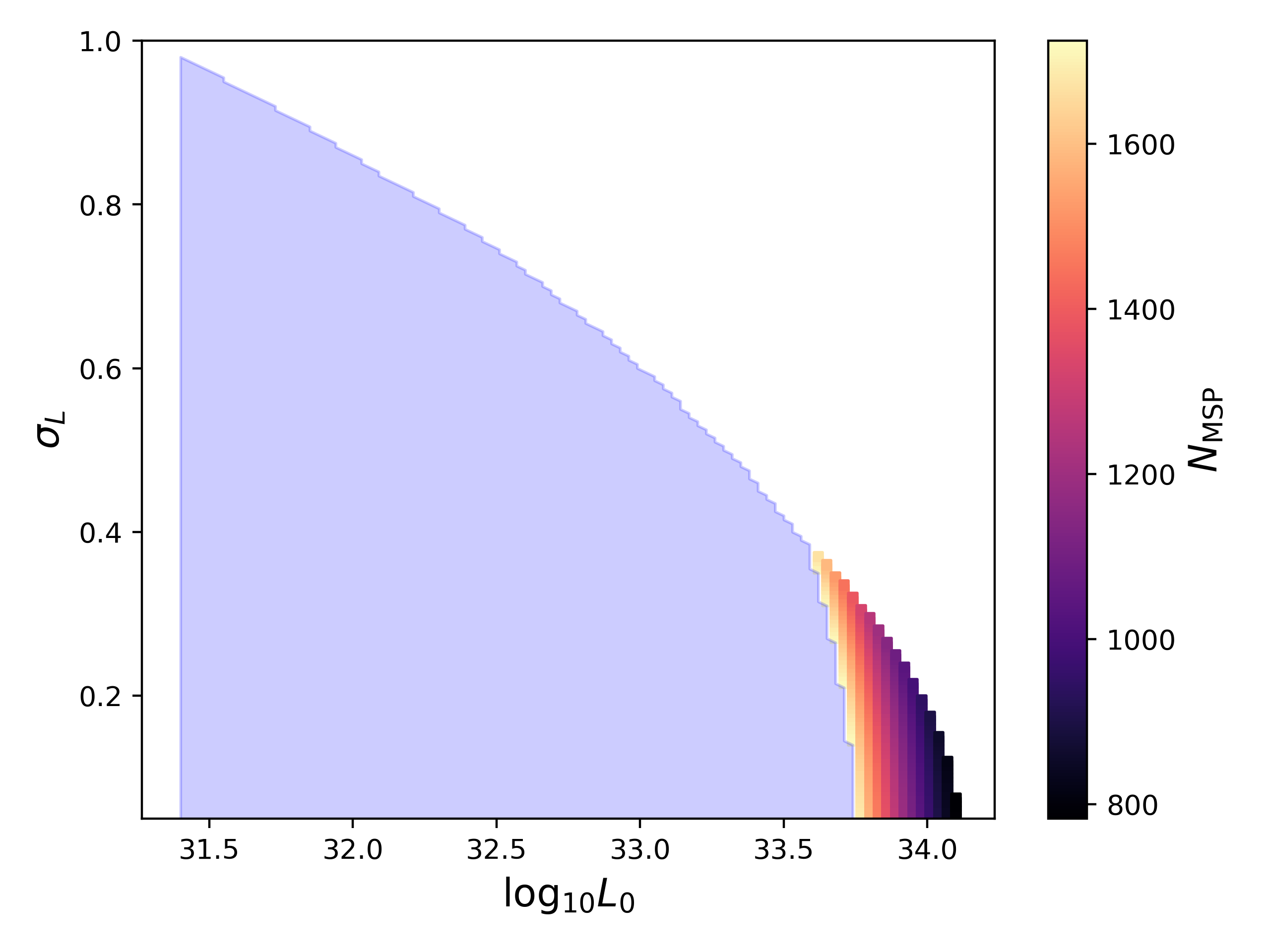}
    \caption{Exclusion regions (light blue) based on the upper limits of the O3 Frequency-Hough all-sky search using the log-normal luminosity function (Eq. \ref{eqn:log-normal}). In these regions, the number of detectable \msps with LIGO/Virgo data exceeds one. We take $B_{\rm int}=150B_{\rm ext}$; $d=8$ kpc, and left: $I_{\rm zz}=10^{38}$kg$\cdot$m$^2$; right: $I_{\rm zz}=5\times10^{38}$kg$\cdot$m$^2$. With these parameters, less than $\mathcal{O}(1\%)$ of the total rotational energy loss of the star is through \gwh radiations.  
    % The boundary of the parameter space is obtained by requiring (1) the \msps above the gamma-ray detection threshold to be fewer than 20\% of all the 4FGL-DR2 sources; and (2) the ratio of the flux emitted by the resolved point sources to be smaller than 20\% of the total flux from the \gc. 
    The upper-right white regions on both plots have been excluded by \fermi.}
    \label{fig:magnetic_strain_model}
\end{figure*}

We present in Fig. \ref{fig:magnetic_strain_model} exclusion regions (light blue) in the $L_0$/$\sigma_L$ parameter space if the neutron-star ellipticity is sustained by an internal magnetic field, as explained in Sec. \ref{mag-def}, and at least one \msp would have been detected in the O3 all-sky search in the \gc. The results are obtained with two values of the moment of inertia, $I_{\rm zz}=10^{38}$ kg$\cdot$m$^2$ (left)  and $I_{\rm zz}=5\times10^{38}$ kg$\cdot$m$^2$ (right), which correspond to different equations of states of \msps. Here, we see that the constraint becomes more stringent if \msps have a higher moment of inertia, given fixed $\Bint$. This is because the ellipticity upper limits from the all-sky search, shown in App. \ref{app:all-sky_ul}, would be lowered by a factor of 5 across all frequencies. Consequently, it leads to a larger probability for \msps to be excluded from existing.

\begin{figure}[htb]
    \centering
    \includegraphics[width=0.49\textwidth]{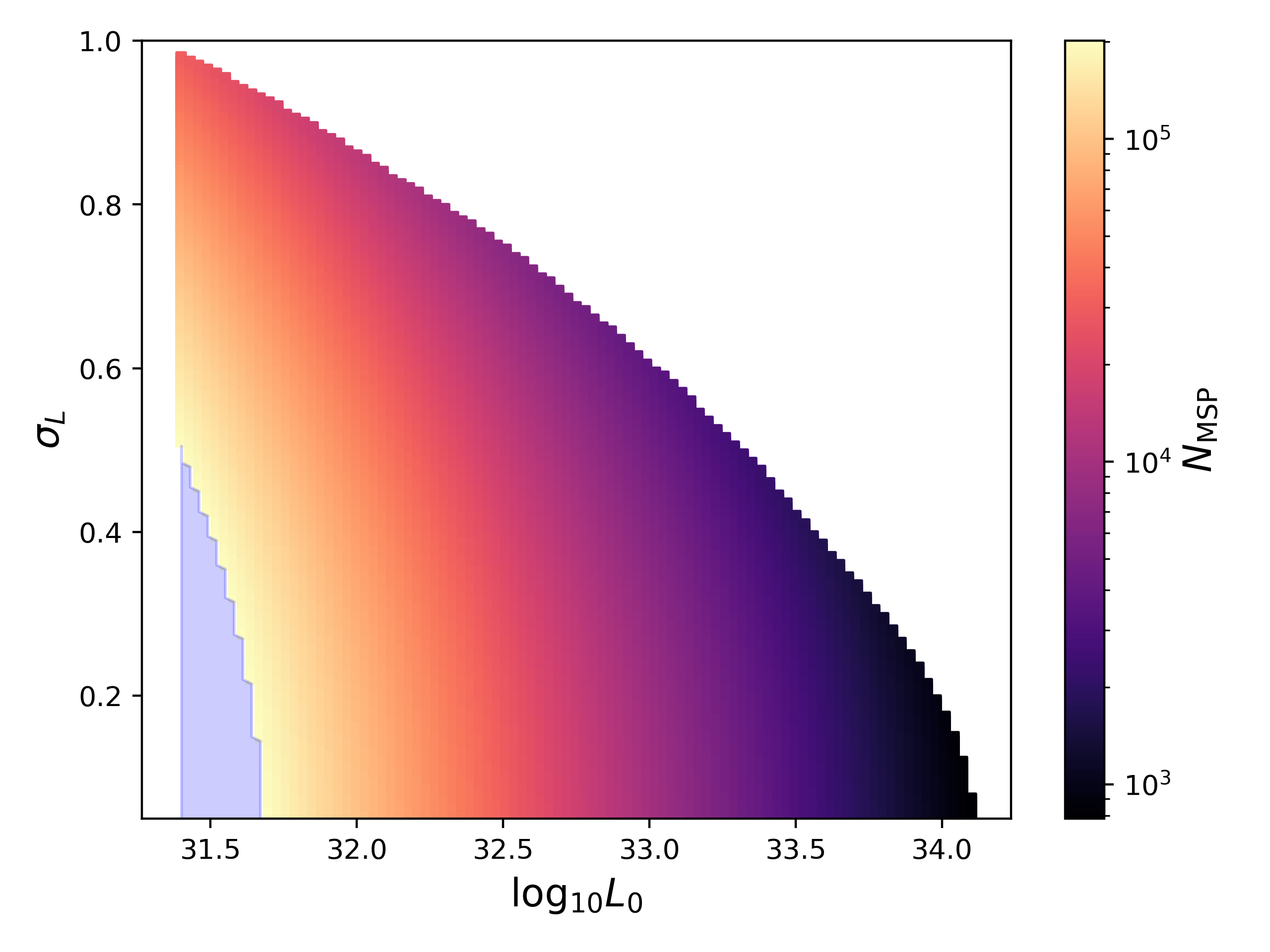}
    \caption{Exclusion regions (light blue) based on the upper limits of the O3 Freuquency-Hough all-sky search using the log-normal luminosity function (Eq. \ref{eqn:log-normal}), employing a probability density function for the ellipticity that assumes 0.5\% of the rotational energy loss of the star goes into \gws. We take $d=8$ kpc and $I_{\rm zz}=10^{38}$kg$\cdot$m$^2$ in this analysis.  
    % The boundary of the parameter space is obtained by requiring (1) the \msps above the gamma-ray detection threshold to be fewer than 20\% of all the 4FGL-DR2 sources; and (2) the ratio of the flux emitted by the resolved point sources to be smaller than 20\% of the total flux from the \gc. 
    The upper-right white region has been excluded by \fermi.}
    \label{fig:0.5perc}
\end{figure}

In Fig. \ref{fig:0.5perc}, we show similar constraints when allowing a fixed fraction of rotational energy to be converted into \gws. The plot corresponds to the fraction taken to be 0.25\% at a fixed moment of inertia. 
% \yz{Star's rotation energy or the lass of the star's rotation energy?} \com{I don't understand the question?}
If \gws take more rotational energy from the star, we are able to exclude larger portions of the $L_0$/$\sigma_L$ parameter space.

% \yz{Just a general question, is there a probability distribution of the $I_{zz}$ for \msps? If there is one, why don't we treat it equally as the ellipticity/frequency distributions?}

\begin{figure*}[htb]
    \centering
    \includegraphics[width=\columnwidth]{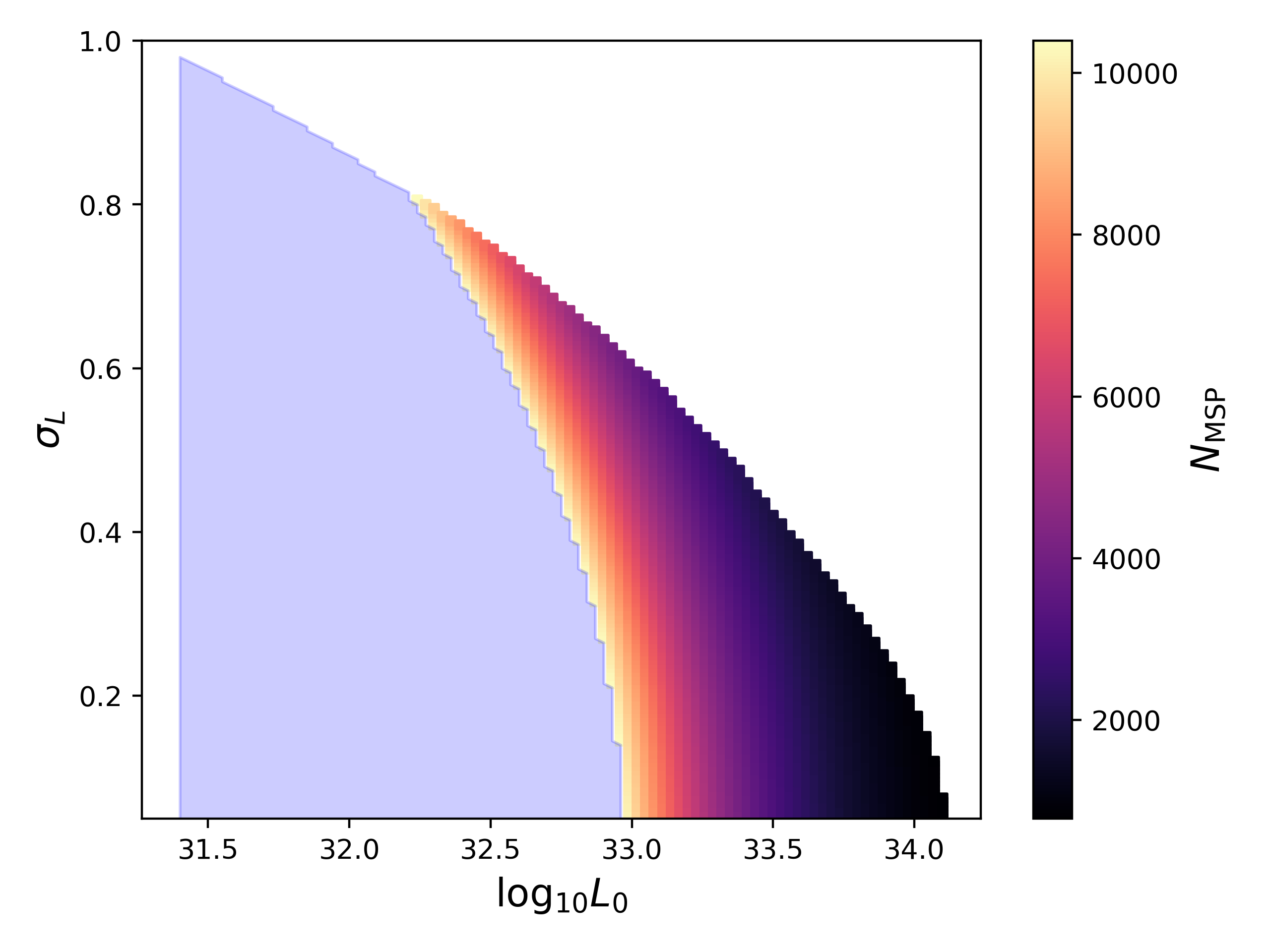}
    \includegraphics[width=\columnwidth]{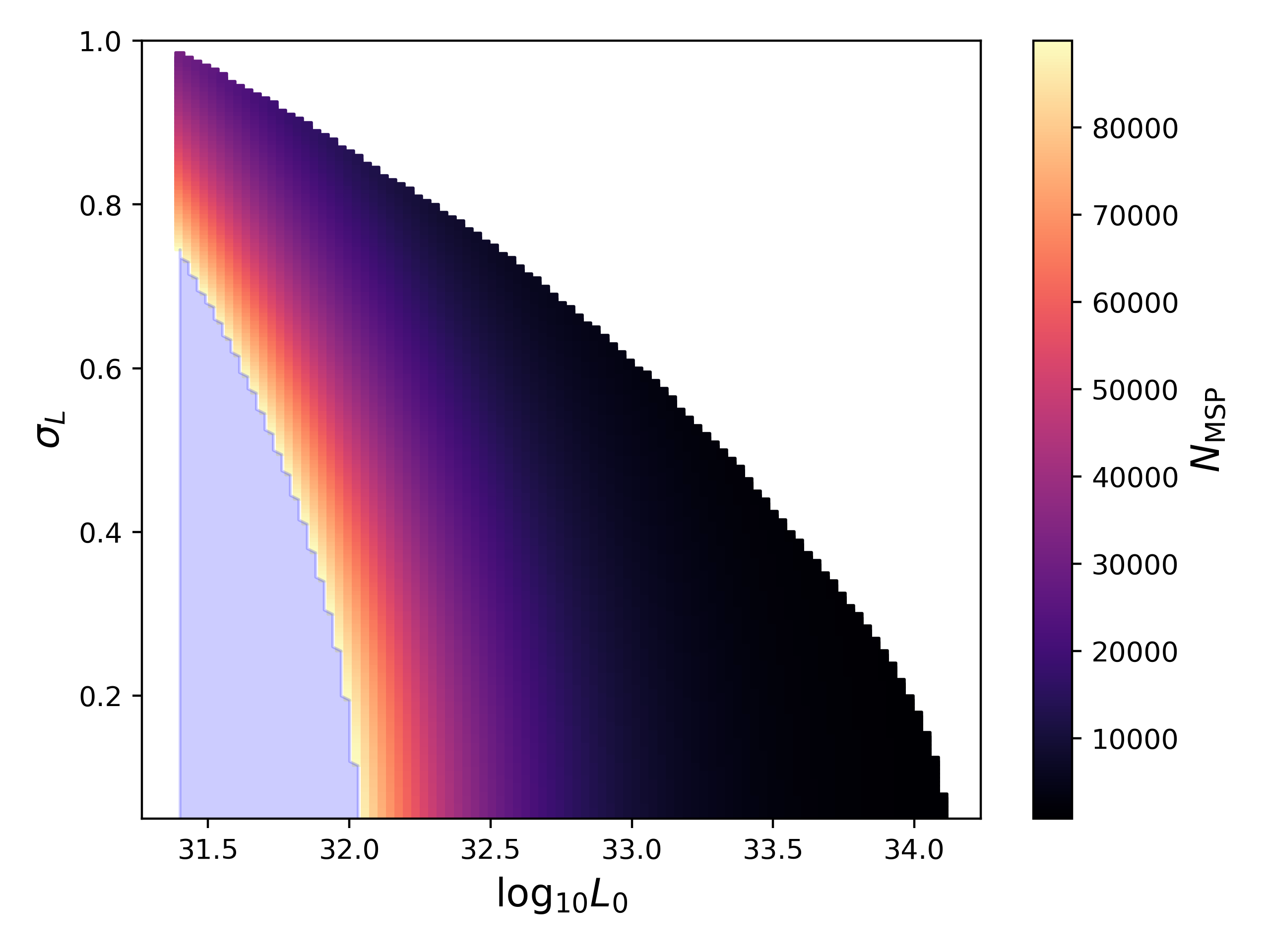}
    \caption{Exclusion regions (light blue) based on the upper limits of the O3 Frequency-Hough all-sky search using the log-normal luminosity function (Eq. \ref{eqn:log-normal}). In these regions, the number of detectable \msps with LIGO/Virgo data exceeds one. We take  $I_{\rm zz}=10^{38}$kg$\cdot$m$^2$ and $d=6$ kpc (left) and $d=8$ kpc (right), and allow only 1\% of rotational energy to be converted to \gws. 
    }
    \label{fig:dist_changes_agnostic}
\end{figure*}

We also determine how sensitive our exclusion regions are to distance reach, picking 6 or 10 kpc to contrast with 8 kpc used throughout the paper. Millisecond pulsars could in fact be closer than 8 kpc to us if they live in the  ``boxy
bulge'' (a part of the Galactic bar) \cite{Macias:2016nev,Bartels:2017vsx} and thus our constraints would be more stringent on them there. We show in Fig. \ref{fig:dist_changes_agnostic} exclusion regions for 6 kpc (left) and 10 kpc (right) systems.

\begin{figure*}[htb]
    \centering
    \includegraphics[width=\columnwidth]{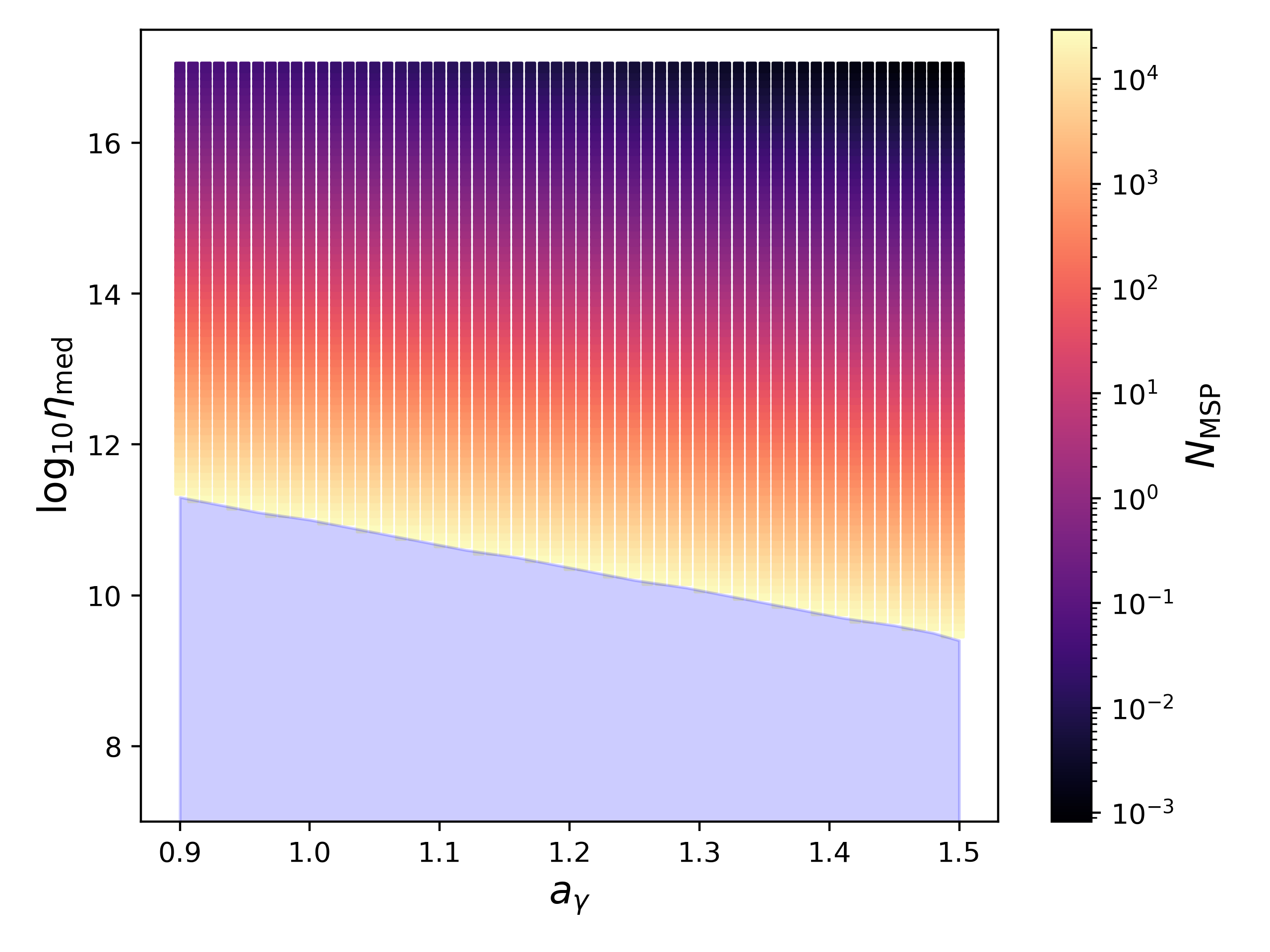}
    \includegraphics[width=\columnwidth]{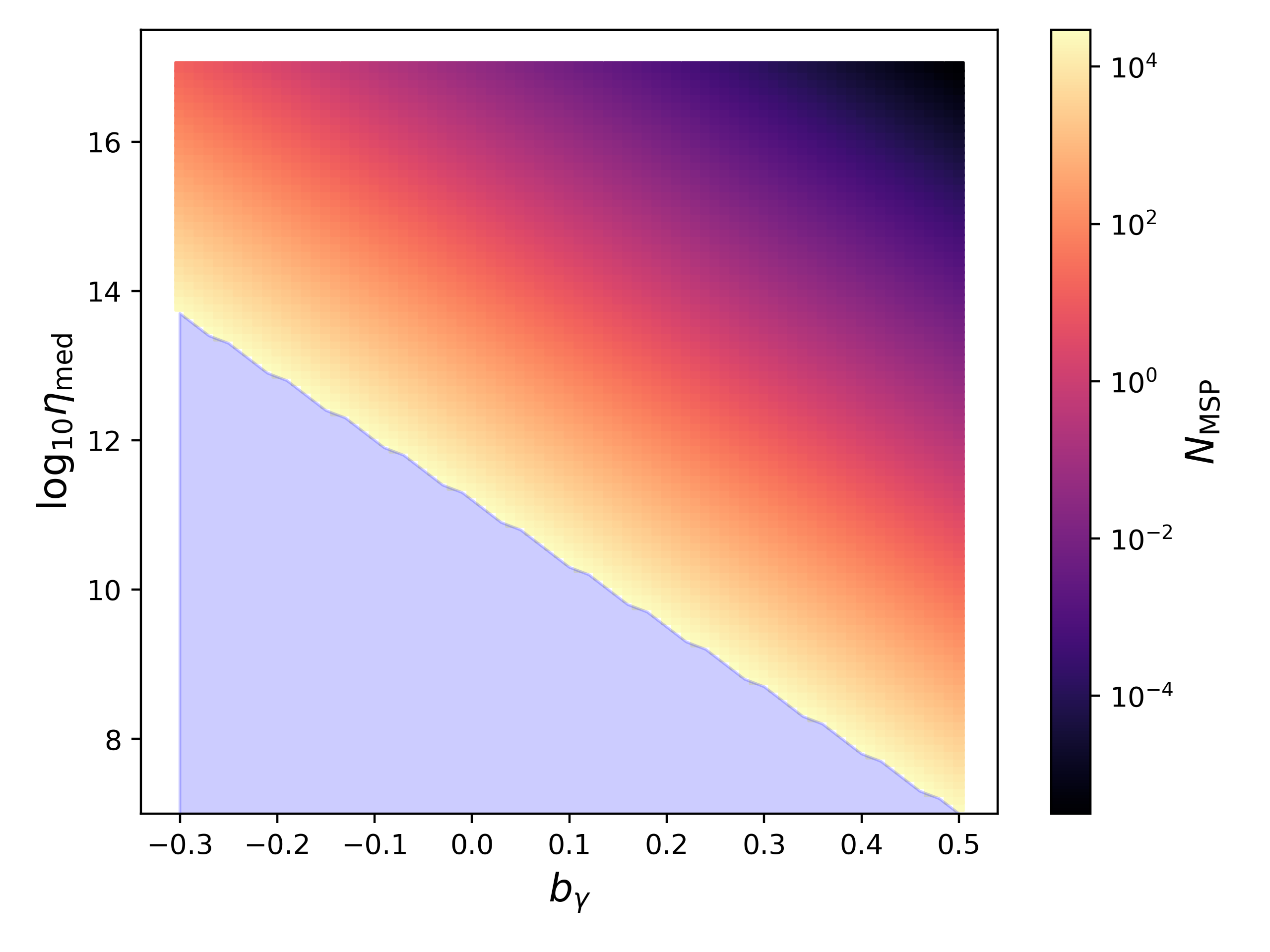}
    \caption{Exclusion regions (light blue) based on the upper limits of the O3 Frequency-Hough all-sky search using the power-law luminosity function (Eq. \ref{eqn:power-law}). In these regions, the number of detectable \msps with LIGO/Virgo data exceeds one. We take  $I_{\rm zz}=10^{38}$kg$\cdot$m$^2$, $d=8$ kpc, and allow only 1\% of rotational energy to be converted to \gws. The range of $N_{\rm MSP}$ is taken to include the values within the error bars of \cite{Ploeg:2020jeh}.
    }
    \label{fig:a-b-gamma}
\end{figure*}

Finally, we also provide exclusion plots for two other parameterizations of the power-law luminosity function model (Eq. \ref{eqn:power-law} in Fig. \ref{fig:a-b-gamma}). We can see that portions of the parameter space for these two choices of power-law luminosity function can also be excluded by continuous-wave upper limits.

\bibliographystyle{apsrev4-1}
\bibliography{biblio}

\end{document}